\documentclass[fleqn,usenatbib]{mnras}

\usepackage[T1]{fontenc}

\DeclareRobustCommand{\VAN}[3]{#2}
\let\VANthebibliography\thebibliography
\def\thebibliography{\DeclareRobustCommand{\VAN}[3]{##3}\VANthebibliography}

\usepackage{graphicx}	% Including figure files
\usepackage{amsmath}	% Advanced maths commands
\usepackage{amssymb}

\title[The Bayesian Global Sky Model (B-GSM)]{The Bayesian Global Sky Model (B-GSM): Validation of a Data Driven Bayesian Simultaneous Component Separation and Calibration Algorithm for EoR Foreground Modelling}   

\author[G. Carter]{
George Carter,$^{1}$\thanks{E-mail: gtc30@cam.ac.uk}
Will Handley,$^{1}$
Mark Ashdown,$^{2}$
and Nima Razavi-Ghods$^{2}$
\\
$^{1}$Institute of Astronomy, University of Cambridge, Cambridge, United Kingdom\\
$^{2}$Cavendish Laboratory, Department of Physics, University of Cambridge, Cambridge, United Kingdom\\
}

\date{Accepted XXX. Received YYY; in original form ZZZ}

\pubyear{\the\year{}}

\begin{document}
\label{firstpage}
\pagerange{\pageref{firstpage}--\pageref{lastpage}}
\maketitle

\begin{abstract}  
We introduce the Bayesian Global Sky Model (B-GSM), a novel data-driven Bayesian approach to modelling radio foregrounds at frequencies <400~MHz. B-GSM aims to address the limitations of previous models by incorporating robust error quantification and calibration. Using nested sampling, we compute Bayesian evidence and posterior distributions for the spectral behaviour and spatial amplitudes of diffuse emission components. Bayesian model comparison is used to determine the optimal number of emission components and their spectral parametrisation. Posterior sky predictions are conditioned on both diffuse emission and absolute temperature datasets, enabling simultaneous component separation and calibration. B-GSM is validated against a synthetic dataset designed to mimic the partial sky coverage, thermal noise, and calibration uncertainties present in real observations of the diffuse sky at low frequencies. B-GSM successfully identifies a model parametrisation with two emission components featuring curved power-law spectra. The Bayesian evidence strongly rejects incorrect models, confirming B-GSM's capability to use the dataset to accurately determine model parametrisation. The posterior sky predictions agree with the true synthetic sky within statistical uncertainty. We find that the root-mean-square (RMS) residuals between the true and posterior predictions for the sky temperature as a function of LST are significantly reduced, when compared to the uncalibrated dataset. This indicates that B-GSM is able to correctly calibrate its posterior sky prediction to the independent absolute temperature dataset. We find that while the spectral parameters and component amplitudes exhibit some sensitivity to prior assumptions, the posterior sky predictions remain robust across a selection of different priors. This is the first of two papers, and is focused on validation of B-GSMs Bayesian framework, the second paper will present results of deployment on real data and introduce the low-frequency sky model which will be available for public download and use.

\end{abstract}


%%%%%%%%%%%%%%%%%%%%%%%%%%%%%%%%%%%%%%%%%%%%%%%%%%

%%%%%%%%%%%%%%%%% BODY OF PAPER %%%%%%%%%%%%%%%%%%

\section{Introduction}
Detection of the cosmological 21cm signal is limited by contamination by bright foreground emission that exceeds the expected signal by 3-6 orders of magnitude~\citep{Pritchard2012,LFSM}. Mitigation of this foreground contamination depends heavily on accurate models of the sky's emission across different angular locations and frequencies. Traditional sky models, such as the Global Sky Model (GSM)~\citep{GSM}, its 2016 update ~\citep{Zheng2016}, and the Low Frequency Sky Model~\citep{LFSM} perform component separation using a Principal Component Analysis (PCA) of the diffuse emission survey maps that form their datasets. 

These models neglect the variability and uncertainty in calibration of these surveys, which is known to be significant \citep{Monsalve2021,Spinelli2021}, and are primarily based on high frequency data from above 1~GHz. Additionally, the PCA based component separation used by these models does not provide any estimate for the uncertainty on the predicted sky. Taken together, these issues limit the usefulness of existing sky models for EoR and 21 cm cosmology, where the frequency range of interest is <200~MHz \citep{Pritchard2012,Liu2013}.
 
In this paper, we present a new data driven Bayesian approach to modelling the low frequency foreground, the Bayesian Global Sky Model (B-GSM). The novel Bayesian framework of B-GSM aims to address the limitations of previous sky models, by introducing robust error quantification and calibration into the model. We use nested sampling~\citep{Skilling2004} to compute Bayesian evidence and to determine posterior distributions for the spectral behaviour and spatial amplitudes of diffuse emission components. Bayesian model comparison~\citep{Trotta2008}, using these Bayesian evidence values, is then used to select the number of emission components and their spectral parametrisation. 

We present the mathematical framework for B-GSM, deriving a joint likelihood function and performing analytical marginalisation. Additionally, we discuss the issue of computational complexity and present an approximation of our marginal likelihood that allows for practical use of our mathematical framework. Finally, we validate the effectiveness of B-GSM by testing it on a synthetic dataset.

\section{Component Separation in B-GSM}
Our model is built on the assumption that the observed sky, \( D \), can be expressed as a function of angular location (\(\Omega\)) and frequency (\(v\)), with an underlying true signal \( D_\mathrm{true}(\Omega,v) \). As in previous sky models, e.g. GSM~\citep{GSM} and LFSM~\citep{LFSM}, we assume that this true signal can be decomposed into the sum of \( k \) components, each described by a spatial map, $M_c(\Omega)$, and a spectrum, $S^c(v)$:
\begin{equation}\label{eq:component_sep}
    D_\mathrm{true}(\Omega,v) = \sum_{c=1}^{k}{M_c(\Omega)S^c(v)}.
\end{equation}

The observed diffuse maps typically have inconsistent and inaccurate calibration of their temperature scale and their temperature zero level. For example, Monsalve et al. 2021 found that the Guzman 45~MHz map \citep{Guzmn2010} required a temperature-scale correction of $1.076 \pm 0.017$ and a zero-level correction of $160 \pm 39$ K in order to fit independent measurements of the sky's absolute temperature \citep{Monsalve2021}. Likewise, the temperature scale of the LWA1 diffuse maps~\citep{LFSM} were found to agree with independent measurements of the sky's absolute temperature at only a $\sim15\%$ level~\citep{Spinelli2021}. We want to correct for calibration errors in our diffuse map dataset as part of our inference. As in \citep{Monsalve2021} we assume that the correction to the calibration for each diffuse map takes the form of a global zero level shift and a global scale factor correction. Specifically, we assume that (for frequency $v$) the correctly calibrated map, $D_{v,\mathrm{cal}}(\Omega)$, is related to the observed map in our dataset, $D_v(\Omega)$, by:
\begin{equation}\label{calibration eq}
    D_{v,\mathrm{cal}}(\Omega) = a_v D_v(\Omega) + b_v.
\end{equation}
where, $a_v$ is the correction to the temperature scale and $b_v$ is the correction to the temperature zero level for the map at frequency $v$. It is these correctly calibrated observed maps that can be related to our true underlying signal. Accounting for the noise in the observed maps, which we denote as \( N(\Omega,v) \), and noting that the calibration correction to the temperature scale also affects the noise, we have that the correctly calibrated sky is related to the true underlying sky as:
\begin{equation}
    D_{v,\mathrm{cal}}(\Omega) = D_\mathrm{true}(\Omega,v) + a_v N(\Omega,v).
\end{equation}
Substituting in our expressions for the calibrated sky (equation \ref{calibration eq}) and the true underlying sky (equation \ref{eq:component_sep}), we find:
\begin{equation}\label{eq:cal_obs_maps}
a_v D_v(\Omega) + b_v = \left(\sum_{c=1}^{k}{M_c(\Omega)S^c(v)}\right)  + a_v N(\Omega,v).
\end{equation}
Which relates the observed map at frequency $v$, $D_v(\Omega)$, to the true underlying sky model, $\sum_{c=1}^{k}{M_c(\Omega)S^c(v)}$, and the noise in the observed sky, $N(\Omega,v)$, via a set of calibration parameters, $a_v$ and $b_v$.
\section{Joint Likelihood}
From the expression in equation (\ref{eq:cal_obs_maps}) we see that constructing a sky model and calibrating our diffuse maps amounts to determining a set of component maps \( M_c \) and spectra \( S^c \), along with calibration parameters $a_v$ \& $b_v$ $\forall$ $v$, given our observed data. This problem is suited to a Bayesian framework, where we infer posterior distributions for the model parameters conditioned on the observed data \citep{Sivia_Skilling}. 

For B-GSM the posterior is conditioned on two independent datasets; a dataset of $n_v$ large area diffuse sky surveys covering the frequency range 45-408~MHz which we will denote as $D$, and a dataset of absolute temperature measurements denoted as $E$. The diffuse maps dataset, $D$, is spatially resolved, however it is poorly calibrated. The absolute temperature dataset, $E$, consists of measurements of the sky absolute temperature as a function of Local Sidereal Time (LST) at equally spaced frequencies between 40~MHz and 200~MHz, at each frequency these measurements cover the full 24 hours of LST. These absolute temperature measurements are not spatially resolved but are correctly and consistently calibrated, they act as a ``ground truth'' and allow inference of the calibration parameters for B-GSM. In the case of real data these absolute temperature measurements would be the $T$ vs LST curves from a radiometer, e.g. from EDGES ~\citep{Mozdzen2017,Mozdzen2018}, or SARAS 2~\citep{Singh2018}. 

Conditioning our posterior on both the (spatially resolved, but poorly calibrated) diffuse maps $D$, and the (spatially unresolved, but well calibrated) absolute temperature measurements $E$, allows B-GSM to perform simultaneous component separation and calibration. In formal mathematical notation, the posterior we want to determine is:
\begin{equation}
    P(a,\Vec{b},M,S|E,D) = \frac{P(E,D| a,\Vec{b},M,S) P(a,\Vec{b},M,S)}{P(E,D)},
\end{equation}
where $P(a,\Vec{b},M,S)$ is the joint prior for our calibration parameters\footnote{Note: we have overloaded our notation $a$ is a $n_v\times n_v$ diagonal matrix of the temperature scale corrections, and $\Vec{b}\in \mathbb{R}^{n_v}$ is a vector of temperature scale corrections. The diagonal elements of $a$ are $a_{v,v}=a_v$ and the elements of $\Vec{b}$ are $\Vec{b}_v=b_v$.} $a$ and $\Vec{b}$, component amplitude maps $M$ and spectral model $S$. The term $P(E,D| a,\Vec{b},M,S)$ is the joint likelihood of observing both our datasets for a given set of component maps $M$, spectral model $S$, and calibration corrections $a$ \& $\Vec{b}$. Recalling that the datasets $D$ and $E$ are independent, and noting that model's sky predictions only depend on the choice of $M$ and $S$, we find that the joint likelihood can be split into two terms~\citep{Sivia_Skilling}:
\begin{equation}\label{joint likelihood definintion}
    P(E, D| a,\Vec{b},M,S) = P(E|M,S)P(D|a,\Vec{b},M,S).
\end{equation}
The first accounts for the likelihood of observing the absolute temperature dataset given our model, $P(E|M,S)$, and the second accounts for the likelihood of observing the dataset formed from the diffuse emission maps $P(D| a,\Vec{b},M,S)$.

\subsection{The Diffuse Likelihood}
To derive an expression for the diffuse likelihood, $P(D| a,\Vec{b},M,S)$, we look to the expression in equation \ref{eq:cal_obs_maps} which relates the (calibrated) observed diffuse maps to the true underlying signal (which we assume can be modelled as the sum of $k$ components) and some additive noise. We will make the simplifying assumption that the noise in the observed maps is uncorrelated between pixels\footnote{In reality, the noise will be correlated between pixels, with the correlated region being of scale equal to the FWHM of the beam for that map. Ignoring these correlations allows our likelihood to be defined on a pixel by pixel basis, this significantly reduces computational complexity.}, allowing us to define a likelihood for a single pixel and take the product of these individual pixel likelihoods~\citep{Stompor2009}:

\begin{equation}
    \label{the diffuse total likelihood term}
    P(D|a,\Vec{b},M,S) = \prod_p P(\Vec{d}_p|a,\Vec{b},\Vec{M}_p,S).
\end{equation}
We have defined the data vector of observed sky temperatures, $\Vec{d}_p \in \mathbb{R}^{n_v}$, (for the $n_v$ frequencies at which we have observations) for pixel $p$. The vector $\Vec{M}_p\in \mathbb{R}^k$ contains the values for each of the $k$ component amplitude maps at pixel $p$. Again we overload our notation, the term $S$ refers to a spectral mixing matrix of shape $n_v\times k$ (this mixing matrix is the same for all pixels). The $c$'th column of the matrix $S$ corresponds to the spectrum of component $c$, $S^c(v)$, evaluated at each of the observed frequencies. Given these definitions, we see that for each pixel the model prediction for the sky vector is given by $S\Vec{M}_p$. Hence, for each pixel, we can relate our observed data vector $\Vec{d}_p$ to our model prediction as:
\begin{equation}
\begin{aligned}
    a\Vec{d}_p +\Vec{b} &= S\Vec{M}_p + a\Vec{n}_p, \\
    \implies \Vec{d}_p &= a^{-1}\left(S\Vec{M}_p - \Vec{b}\right) + \Vec{n}_p. 
\end{aligned}
\end{equation}
The vector $\Vec{n}_p$ is the additive noise present in the observations of the sky at pixel $p$. If we assume that this noise is Gaussian, then we can express our log likelihood as:

\begin{multline}\label{pix by pix complicated}
    2\ln{P(\Vec{d}_p|a,\Vec{b},\Vec{M}_p,S)} = -\left[\Vec{d}_p - a^{-1}\left(S\Vec{M}_p -\Vec{b}\right)\right]^T N_p^{-1}\\ \left[\Vec{d}_p - a^{-1}\left(S\Vec{M}_p -\Vec{b}\right)\right] - \ln\left(\left|2\pi N_p \right|\right).
\end{multline}
Note that $N_p\in \mathbb{R}^{n_v\times n_v}$ is the noise covariance matrix for pixel $p$, which we assume to be diagonal (i.e. noise between maps is uncorrelated). This expression can be simplified, let us define the map independent terms as:

\begin{equation}
\label{notation definition}
\begin{aligned}
    &C_p^{-1} = S^T a^{-1}N_p^{-1}a^{-1} S \in \mathbb{R}^{k\times k},\\
    &\Vec{\Gamma}_p = S^T a^{-1}N_p^{-1}a^{-1} \left(a \Vec{d}_p +\Vec{b} \right) \in \mathbb{R}^{k}, \\
    &g_p = (a\Vec{d}_p + \Vec{b})^T a^{-1}N_p^{-1}a^{-1} (a\Vec{d}_p + \Vec{b}) + \ln{\left(\left|2\pi  N_p \right|\right)} \in \mathbb{R}.
\end{aligned}
\end{equation}
Here, the matrix $C_p$ can be thought of as the noise covariance matrix projected onto the set of components. Using this notation, equation (\ref{pix by pix complicated}) can then be written as:
\begin{equation}
\label{single pix diffuse term}
    2\ln{P\left(\Vec{d}_p|a,\Vec{b},\Vec{M}_p,S\right)} = -\Vec{M}_p^T C_p^{-1} \Vec{M}_p + 2\Vec{\Gamma}_p^T\Vec{M}_p - g_p,
\end{equation}
which describes the likelihood of observing the data vector, $\Vec{d}_p$, for a specific set of calibration parameters, $a$ and $\Vec{b}$, component maps (at pixel $p$) $\Vec{M}_p$ and spectral model $S$. This term only accounts for the diffuse dataset, we must also derive a term to account for the absolute temperature dataset.

\subsection{The Absolute Temperature Likelihood}
The absolute temperature term, $P(E|M,S)$, compares our sky model's predictions (for a specific $M$, $S$) to the absolute temperature dataset. This dataset, $E$, takes the form of a series of temperature vs LST ($T$ vs LST) curves for frequencies between 40 and 200~MHz. To compare our model to this dataset, we must produce a set of simulated antenna temperatures for our model, $T_{\mathrm{mod},v,\mathrm{LST}}(M,S)$, at every frequency and LST in the absolute temperature dataset. Producing these simulated antenna temperatures for our model (at a specific frequency and LST) amounts to convolving our model's predicted sky (for a given $M$ and $S$) at that frequency with an antenna beam model (for that frequency) rotated to the correct LST, 

\begin{equation}
\label{simulated EDGES temp at LST and v}
    T_{\mathrm{mod},v,\mathrm{LST}}(M,S) = \Vec{B}_{v,\mathrm{LST}}\cdot \Vec{D}_{\mathrm{true}}(v,M,S).
\end{equation}
Here, $\Vec{D}_{\mathrm{true}}(v,M,S) \in \mathbb{R}^{n_p}$ is the model's predicted sky map at frequency $v$ ($n_p$ is the number of pixels in the map), and $\Vec{B}_{v,\mathrm{LST}}\in \mathbb{R}^{n_p}$ is our beam model for that frequency and rotated to the correct LST. Assuming Gaussian noise on the absolute temperature observations, we obtain a likelihood of:

\begin{multline}
    \label{EDGES likelihood term}
    2\ln{P\left(E|M,S\right)} = -\sum_\mathrm{LST}\sum_v \left(\frac{T_{E,v,\mathrm{LST}} - T_{\mathrm{mod},v,\mathrm{LST}}(M,S)}{\sigma_{E,v,\mathrm{LST}}}\right)^2 \\- \sum_\mathrm{LST}\sum_v\ln{\left(2\pi\sigma_{E,v,\mathrm{LST}}^2\right)}.
\end{multline}
Where $T_{E,v,\mathrm{LST}}$ is the observed antenna temperature (at a given frequency and LST), $T_{\mathrm{mod},v,\mathrm{LST}}(M,S)$ is the simulated antenna temperature for our model (for a given $M$ and $S$), and $\sigma_{E,v,\mathrm{LST}}$ is the noise on the observation. The double sum runs over all observed frequencies, $v$, and LST's.

\subsection{The Joint Likelihood}
The joint likelihood is the product of the diffuse and absolute temperature likelihoods, the joint log-likelihood is therefore:
\begin{equation}
\label{final joint likelihood with calibration}
\begin{aligned}
    2\ln P\left(E,D|a,\Vec{b},M,S\right) = - \sum_\mathrm{LST}\sum_v\ln{\left(2\pi\sigma_{E,v,\mathrm{LST}}^2\right)}&\\-\sum_\mathrm{LST}\sum_v \left(\frac{T_{E, v,\mathrm{LST}} - T_{\mathrm{mod},v,\mathrm{LST}}(M,S)}{\sigma_{E,v,\mathrm{LST}}}\right)^2& \\ 
    -\sum_p\left(\left[\Vec{M}_p - C_p\Gamma_p\right]^T C_p^{-1} \left[\Vec{M}_p - C_p\Gamma_p\right] - \Gamma_p^T C_p \Gamma_p +g_p\right)&.\\
\end{aligned}
\end{equation}
Note that we have used completing the square to rearrange the single pixel diffuse likelihood (equation \ref{single pix diffuse term}).

\section{Block Matrix Formalism}
So far, we have worked using a pixel-by-pixel approach to define our likelihoods. However, if we look at the absolute temperature likelihood (equation \ref{EDGES likelihood term}) we see that the simulated antenna temperatures for our model $T_{\mathrm{mod},v,\mathrm{LST}}$ (equation \ref{simulated EDGES temp at LST and v}) are defined by a convolution across the whole predicted sky at each frequency $v$. As such, it makes sense to define the likelihood on a map-by-map basis. To do this we use block matrices; let $\Vec{M}$ be the block vector with blocks $\Vec{M}_p$:
\begin{align}
    \Vec{M} = \begin{bmatrix} \Vec{M}_1 \\ \Vec{M}_2 \\ \vdots \\ \Vec{M}_{n_p} \end{bmatrix} \in \mathbb{R}^{(k\cdot n_p)},
\end{align}
and let $\Vec{\Gamma}$ be the block vector with blocks $\Vec{\Gamma}_p$,
\begin{align}
    \Vec{\Gamma} = \begin{bmatrix} \Vec{\Gamma}_1 \\ \Vec{\Gamma}_2 \\ \vdots \\ \Vec{\Gamma}_{n_p} \end{bmatrix} \in \mathbb{R}^{(k\cdot n_p)},
\end{align}
and let $C^{-1}$ be the block diagonal matrix with blocks $C_p^{-1}$,
\begin{align}
    C^{-1} = \begin{bmatrix} C_1^{-1} & 0 & \ldots & 0 \\ 0 & C_2^{-1} &\ldots &0 \\ \vdots &\vdots &\ddots &\vdots\\ 0 &0 &\ldots &C_{n_p}^{-1}\end{bmatrix} \in \mathbb{R}^{(k\cdot n_p)\times (k\cdot n_p)},
\end{align}
then we can express the diffuse likelihood as:
\begin{equation}\label{diffuse likelihood as a matrix equation}
    2\ln P(D|a,\Vec{b},\Vec{M},S) = -\Vec{M}^T C^{-1} \Vec{M} +2\Vec{\Gamma}^T\Vec{M} -g,
\end{equation}
where $g = \sum_p g_p$, and all other terms have their previously stated definitions.

Similarly, for the absolute temperature likelihood, let us begin by defining a (block diagonal) spectral mixing matrix that produces a realisation of our model's sky prediction at a single frequency, $v$, let $A(S,v) \in \mathbb{R}^{n_p\times(k\cdot n_p)}$ be:
\begin{align}
    A(S,v) = \begin{bmatrix} \begin{bmatrix}S_1(v) & \ldots &S_k(v)\end{bmatrix} & \ldots &0 \\ \vdots &\ddots &\vdots \\ 0 &\ldots & \begin{bmatrix}S_1(v) & \ldots &S_k(v)\end{bmatrix} \end{bmatrix}.
\end{align}
The block on the diagonal, $\begin{bmatrix}S_1(v) & \ldots &S_k(v)\end{bmatrix} \in \mathbb{R}^{1\times k}$, is a row vector of the $k$ component spectra evaluated at the frequency $v$. Going further with this block matrix formalism, we define:
\begin{align}
    A(S) = \begin{bmatrix} A(S,v_1) \\ A(S,v_2) \\ \vdots \\ A(S,v_{n_E})\end{bmatrix} \in \mathbb{R}^{(n_p\cdot n_E) \times (k\cdot n_p)},
\end{align}
the block matrix formed from the mixing matrices, $A(S,v)$. Where each block is evaluated at a different frequency $v_i\in\{v_1,\ldots,v_{n_E}\}$, where $\{v_1,\ldots,v_{n_E}\}$ is the set of frequencies at which there are absolute temperature observations. From this we get that the vector of predictions of the sky at all the frequencies (at which there are absolute temperature observations) is:
\begin{align}
    \Vec{D}_\mathrm{true}(M,S) = A(S) \Vec{M} = \begin{bmatrix} \Vec{D}_\mathrm{true}(v_1,M,S) \\ \vdots \\ \Vec{D}_\mathrm{true}(v_{n_E},M,S)\end{bmatrix} \in \mathbb{R}^{(n_p \cdot n_E)}.
\end{align}
Now let,
\begin{align}
    B(v) = \begin{bmatrix} \Vec{B}(v)_{\mathrm{LST_1}} \ldots \Vec{B}(v)_{\mathrm{LST_{final}}}\end{bmatrix} \in \mathbb{R}^{n_p\times n_\mathrm{LST}},
\end{align}
be the set of beam models at frequency $v$ and at all the observed LSTs with each, $\Vec{B}_{v,\mathrm{LST_i}}\in\mathbb{R}^{n_p}$, being this beam model rotated to $\mathrm{LST_i} \in \{\mathrm{LST_1},\ldots,\mathrm{LST_{final}}\}$. Finally, if we let the full set of beams at all frequencies and LSTs be given by the rectangular block diagonal matrix,
\begin{align}
    B = \begin{bmatrix} B(v_1) & 0 & \ldots & 0 \\ 0 & B(v_2) &\ldots & 0 \\ \vdots &\vdots &\ddots &\vdots \\ 0 & 0 &\ldots &B(v_{n_E}) \end{bmatrix} \in \mathbb{R} ^ {(n_p \cdot n_E) \times (n_\mathrm{LST} \cdot n_E)},
\end{align}
then the vector of simulated antenna temperatures (for our model) at every observed frequency and LST is,
\begin{equation}
    \Vec{T}_{\mathrm{mod}}(M,S) = B^T \Vec{D}_\mathrm{true}(M,S) = B^T A(S) \Vec{M} \in \mathbb{R} ^ {(n_\mathrm{LST} \cdot n_E)}.
\end{equation}
defining the following map independent terms:
\begin{equation}
\label{notation definition EDGES}
\begin{aligned}
    &C_E^{-1} = A(S)^T B N_E^{-1} B^T A(S) \in \mathbb{R}^{(k\cdot n_p)\times (k\cdot n_p)},\\
    &\Vec{Z}_E = A(S)^T B N_E^{-1} \Vec{T}_E \in \mathbb{R}^{(k\cdot n_p)}.
\end{aligned}
\end{equation}
where $N_E$ is the diagonal noise covariance matrix for the observed sky absolute temperatures, we obtain the following expression for the absolute temperature likelihood:
\begin{multline}\label{EDGES likelihood as a matrix expression}
    2\ln P\left(E|\Vec{M},S\right) = -\Vec{M} C_E^{-1}\Vec{M} + 2\Vec{Z}_E^T \Vec{M} \\- \Vec{T}_E^T N_E^{-1} \Vec{T}_E - \ln\left(\left|2\pi N_E\right|\right),
\end{multline}
where $\Vec{T}_E\in\mathbb{R}^{n_E\cdot n_\mathrm{LST}}$ is the vector of the absolute temperature observations at all $n_\mathrm{LST}$ LSTs and all $n_E$ frequencies.

The joint likelihood is the product of the diffuse likelihood and the absolute temperature likelihood. Using the block matrix definitions for each of these (equations \ref{diffuse likelihood as a matrix equation} and \ref{EDGES likelihood as a matrix expression} respectively) we can write the joint log-likelihood as:
\begin{equation}\label{joint likelihood as a matrix expression}
\begin{aligned}
    2\ln P(E,D|a,\Vec{b},\Vec{M},S) = -\Vec{M}^T\left(C^{-1} + C_E^{-1} \right) \Vec{M} \\ +2 \left(\Vec{\Gamma}+\Vec{Z}_E\right)^T \Vec{M} \\- \Vec{T}_E^T N_E^{-1} \Vec{T}_E - \ln\left(\left|2\pi N_E\right|\right) - g,
\end{aligned}
\end{equation}
This is the joint likelihood of observing both our datasets, $D$ and $E$, for a specific set of calibration parameters, $a$ and $\Vec{b}$, component amplitude maps $\Vec{M}$, and spectral parameters, $S$. We see that in this block matrix formalism, the joint likelihood can be written as a single Gaussian function with a covariance matrix given by $\left(C^{-1} + C_E^{-1} \right)^{-1}$. Note that this expression for the joint likelihood is entirely equivalent to the earlier expression in equation (\ref{final joint likelihood with calibration}), we have simply rearranged the expression.

\subsection{Analytical Marginalisation}
Direct sampling of the full posterior is computationally prohibitive due to high dimensionality. Instead, we perform analytic marginalisation over the component maps, reducing the problem to sampling the marginal posterior of calibration and spectral parameters. Assuming that, a priori, the component amplitude maps do not depend on the calibration parameters, the marginal likelihood is expressed as:

\begin{equation}\label{marginalisation integral}
    P(E,D|a,\Vec{b},S)) = \int{P(E,D|a,\Vec{b},\Vec{M},S)P(\Vec{M}|S)\mathrm{d}\Vec{M}}.              
\end{equation}  

To evaluate this integral, and determine the marginal likelihood function, we must choose a prior for the component map amplitudes. To ensure that our marginal likelihood has the same functional form as the joint likelihood, we will choose a conjugate prior for the component map amplitudes. Thus,  we will choose a Gaussian prior for the component maps: 
\begin{equation}
    P(\Vec{M}|S)=N\left(\Vec{\mu}_0(S),C_0(S)\right).
\end{equation}
Where the term $\Vec{\mu}_0(S) \in \mathbb{R}^{(k\cdot n_p)}$ is the prior mean set of component maps and $C_0(S) \in \mathbb{R}^{(k\cdot n_p)\times (k\cdot n_p)}$ is the prior covariance matrix. In general, both $\Vec{\mu}_0(S)$ and $C_0(S)$ will be functions that depend on the spectral parameters, $S$. In this study we choose to use a mean of zero, though a Gaussian prior with non-zero mean may be worth investigating in future work. Using this Gaussian prior, and the joint likelihood (equation \ref{joint likelihood as a matrix expression}), we can analytically solve the marginalisation integral (equation \ref{marginalisation integral}) to find the marginal likelihood as:
\begin{equation}
\label{marginal likelihood fully correct}
\begin{aligned}
    2\ln(P(E,D|a,\Vec{b},S)) = -\ln\left(\left|C_0(S)\right|\left|C^{-1} +C_E^{-1} + C_0(S)^{-1}\right|\right)\\- \Vec{T}_E^T N_E^{-1} \Vec{T}_E - \ln\left(\left|2\pi N_E\right|\right) - g \\ + \left(\Vec{\Gamma}+\Vec{Z}_E\right)^T\left(C^{-1} + C_E^{-1} + C_0(S)^{-1}\right)^{-1}\left(\Vec{\Gamma}+\Vec{Z}_E\right).
\end{aligned}
\end{equation}
Where all terms have their previous definitions in terms of block matrices and block vectors.

\section{Approximate Marginalisation}\label{approximate marginalisation}
The expression that we have just derived (equation \ref{marginal likelihood fully correct}) is the marginal likelihood of observing our two datasets for a specific calibration and spectral model, irrespective of the component map amplitudes. Whilst this expression does mean we only need sample a limited number of parameters, it does have limitations. Instead of defining our likelihood as the product of single pixel likelihoods each with $k$ dimensional vectors, we have had to work on a map by map basis with $k\cdot n_p$ dimensional vectors. This is due to the fact that the simulated antenna temperatures, $T_{\mathrm{mod},v,LST}$, are defined by a convolution over the whole sky. The size of these vectors and matrices will pose a significant challenge to sampling the posterior, using this likelihood. This is due to the fact that the computational complexity of matrix inversion (at best) scales as $\mathcal{O}(n^{2.373})$ (here $n$ is matrix size)~\citep{Davie2013}. The marginal likelihood we have derived (equation \ref{marginal likelihood fully correct}) requires computing the inverse and determinant for a dense matrix (the $C_E$ matrix) of size $(k\cdot n_p)\times (k\cdot n_p)$. Where $n_p$ is the number of pixels in the component amplitude maps for our model, and $k$ is the number of components for our model. To generate posterior samples, we will need to perform these matrix operations millions of times. We see that for any reasonable map resolution, such sampling will be prohibitively expensive.

To avoid this $\mathcal{O}(n^{2.373})$ to $\mathcal{O}(n^{3})$ growth in computational complexity, we derive an approximate pixel-by-pixel likelihood. To derive this approximate likelihood, we begin by noting that our full joint posterior can be written as:
\begin{equation}
    P(a,\Vec{b},\Vec{M},S|E,D) =\frac{P(E| \Vec{M},S)}{P(E)} \times P(a,\Vec{b},\vec{M},S|D),
\end{equation}
i.e. we split the analysis into two steps. First conditioning the posterior on the diffuse dataset, and then using this diffuse posterior as the prior for the analysis of the absolute temperature dataset~\citep{Sivia_Skilling}. Using the product rule for conditional probabilities gives,
\begin{multline}\label{rewritten joint posterior}
    P(a,b,\Vec{M},S|E,D) = \frac{P(E| \Vec{M},S)}{P(E)} \\ \times P(\Vec{M}|a,\Vec{b},S, D) P(a,\Vec{b},S|D).
\end{multline}
Substituting this expression for the full joint posterior into our marginalisation integral, we find:
\begin{multline}\label{marginalisation integral rewritten}
    P(a,\Vec{b},S|E,D) = \frac{P(D|a,\Vec{b},S)P(a,\Vec{b},S)}{P(E,D)} \\ \times \int P(E|\Vec{M},S) \times P(\Vec{M}|a,\Vec{b},S, D) \mathrm{d}\Vec{M}.
\end{multline}
Where $P(\Vec{M}|a,\Vec{b},S, D)$ is the conditional distribution of component amplitudes, conditioned on a specific set of model parameters, $a$, $\Vec{b}$ and $S$, and the diffuse dataset $D$. Note that this distribution is only conditioned on the diffuse dataset, and does not include the absolute temperature data. We can show that this distribution is Gaussian with a mean given by:
\begin{equation}\label{mean maps}
    \left<\Vec{M}|a,\Vec{b},S,D\right> = (C^{-1} + C_0(S)^{-1})^{-1}\Vec{\Gamma},
\end{equation}
with a covariance matrix given by $(C^{-1} + C_0(S)^{-1})^{-1}$. If we make the assumption that this conditional distribution is very narrow around its mean value, then we may roughly approximate it as a delta function:
\begin{equation}
    P(\Vec{M}|a,\Vec{b},S, D) \approx \delta\left(\Vec{M} - \left<\Vec{M}|a,\Vec{b},S,D\right>\right).
\end{equation}
Substituting this approximation into the integral in equation (\ref{marginalisation integral rewritten}) we obtain the following approximation of the marginal posterior (conditioned on both datasets):
\begin{multline}\label{approximatedion}
    P(a,\Vec{b},S|E,D) \approx\frac{P(D|a,\Vec{b},S)P(a,\Vec{b},S)}{P(E,D)} \\ \times P(E|\left<\Vec{M}|a,\Vec{b},S,D\right>,S).
\end{multline}
If we choose our component amplitude prior, $P(\Vec{M}|S)$, to be defined pixel-by-pixel with a block diagonal covariance matrix with $n_p$ identical $k\times k$ blocks $c_0$, i.e. 
\begin{equation}
    P(\Vec{M}_p|S)=N(\Vec{0},c_0(S)) \quad \forall \quad p \in \{1,\ldots,n_p\}.
\end{equation}
Then this approximate likelihood does not require the inversion of any large dense matrices, and can be rewritten as the sum across all pixels as:

\begin{multline}
\label{approximation logL}
    2\ln{P(E,D| a,\Vec{b},S)} \approx -\left<\Vec{M}|a,\Vec{b},S,D\right>^T C_E^{-1} \left<\Vec{M}|a,\Vec{b},S,D\right> \\+ 2\Vec{Z}_E^T \left<\Vec{M}|a,\Vec{b},S,D\right> - \Vec{T}_E^T N_E^{-1}\Vec{T}_E - \ln(|2\pi N_E|) \\+\sum_p \left( \Vec{\Gamma}_p^T\left(C_p^{-1} + c_0(S)^{-1}\right)^{-1}\Vec{\Gamma}_p\right) \\+\sum_p \left(-\ln\left(\left|c_0(S)\right|\left|C_p^{-1} + c_0(S)^{-1}\right|\right) - g_p \right).
\end{multline}

\subsection{Component Amplitudes}
After generating a set of samples from the marginal posterior, $\{a,\Vec{b},S\}_\mathrm{posterior}$, we must also generate a set of posterior samples for the component amplitude maps, $\{M\}_\mathrm{posterior}$. The $i$-th posterior sample amplitude maps, $M_i$, is drawn from the conditional distribution, $M_i \sim P(M|a_i,\Vec{b}_i,S_i,D)$, conditioned on the $i$-th sample from the marginal posterior $a_i,\Vec{b}_i,S_i$ and the dataset, $D$. The $i$-th sample component maps are generated pixel by pixel from the distribution $P(\Vec{M}_p|a,\Vec{b},S,\Vec{d}_p)$.
The weighting of these samples comes from the weights calculated during sampling of the approximate marginal posterior.

\subsection{Interpretation of the Approximation}
From equation \ref{approximatedion} we see that the marginal posterior samples correspond to samples from the distribution $P(a,\Vec{b},S|D)$, that are re-weighted in light of the absolute temperature observations. The posterior samples generated for the component amplitudes come from the analytically defined distribution $P(\Vec{M}|a,\Vec{b},S,D)$. With each posterior sample of component amplitudes being conditioned on a corresponding marginal posterior sample, $a,\Vec{b},S$. This means that the joint posterior samples we generate, $\{a_i,\Vec{b}_i,\Vec{M}_i,S_i\}_\mathrm{posterior}$, are actually samples drawn from the joint posterior distribution conditioned on the diffuse maps, $P(a,\Vec{b},\vec{M},S|D)$. Which are then re-weighted to account for the absolute temperature observations in the dataset, $E$. This re-weighing is based on evaluating the absolute temperature likelihood for the spectral model $S$ and the mean diffuse posterior maps $\left<\Vec{M}|a,\Vec{b},S,D\right> = \left(C^{-1}+C_0^{-1}\right)^{-1}\Vec{\Gamma}$.

From this interpretation, it is apparent that, the spatial structure of the predicted sky is informed by the diffuse maps during the sampling from the $P(a,\Vec{b},\Vec{M},S|D)$ distribution, and that the reweighting of these samples then introduces the information about the spectral behaviour and calibration from the absolute temperature observations. Intuitively, this makes sense. We know that, the majority of the information about the spatial structure is contained in the spatially resolved (but poorly calibrated) diffuse maps. Whereas, the majority of the spectral information is contained in the well calibrated (but 1d, not spatially resolved) $T$ vs LST curves.

\section{Degeneracy}
In the earlier GSM \citep{GSM} and its 2016 update \citep{Zheng2016}, the component separation formalism contains a fundamental degeneracy. The degeneracy takes the form of an invertible $k\times k$ matrix $X$. For any given solution $S$ and $\Vec{M}_p$, there exists an equivalent solution with mixing matrix $\Tilde{S}=SX$ and component map vector $\Tilde{\Vec{M}}_p=X^{-1}\Vec{M}_p$. That is to say, for any solution we can take a linear combination of the component spectra and, for a reciprocal combination of component amplitude maps, obtain an identical prediction for the sky.

In B-GSM this degeneracy is broken when we explicitly select a particular functional form for our component spectra $S^c(v)$. However, in regions of parameter space where two or more spectra become linearly dependent, then it is possible to take a linear combination of the spectra and obtain a new mixing matrix. In such areas of parameter space, the solution for the associated set of component maps $\Vec{M}$ is degenerate.

\subsection{A Conditional Prior for the Component Amplitudes}\label{conditional marginalisation}
A solution to this degeneracy lies in our choice of prior for the component amplitudes. Looking at the definition of the mean component amplitudes (equation \ref{mean maps}) and the approximate likelihood (equation \ref{approximation logL}) we see that the covariance matrix of the map prior, $c_0(S)$, acts as a regularisation term. Given that the $c_0(S)$ matrix has a dependence on $S$, the degree of regularisation can be made to respond adaptively to the linear dependence of the spectra. We can think of this adaptive regularisation as requiering the ``volume'' of the allowed space of compoent amplitudes to expand or contract in responce to the linear dependence of the columns of the matrix $S$. 

It can be shown that the prior distribution of the sky temperatures is related to the prior distribution of the component amplitudes via the spectral mixing matrix, $S$. As such if we define a prior covariance for the sky temperatures, $C_\mathrm{Sky}\in\mathbb{R}^{n_v\times n_v}$, it is possible to define a conditional prior covariance for the component amplitudes, $c_0(S)$, as a function of $S$ and $C_\mathrm{Sky}$. Using the Moore-Penrose pseudo-inverse of $S^T$~\citep{matrix_cookbook}, and the relation between the sky temperature and component amplitude covariance matrices, we obtain the following expression for $c_0(S)$: 
\begin{equation}
    c_0(S) = \left( S^T C_\mathrm{Sky}^{-1} S \right)^{-1}.
\end{equation}
This definition of the $c_0(S)$ matrix will provide a broad prior in regions where $S$ is not linearly dependent, and an increasingly restrictive prior as $S$ becomes closer to being linearly dependent. Additionally, this definition of $c_0(S)$ means that the task of choosing a prior covariance matrix for the component amplitudes (dificult as a priori we do not know the emission components) is reduced to the task of defining a prior covariance for the sky temperatures $C_\mathrm{Sky}$. This prior sky covariance, $C_\mathrm{Sky}$, encodes our assumptions about the covariance of the sky temperatures\footnote{Note that in effect we have defined a Gaussian prior for the sky amplitudes $P(\mathrm{Sky})=N(\mu_\mathrm{Sky},C_\mathrm{Sky})$ and then projected this sky prior on to the component map amplitudes. We have chosen $\Vec{\mu}_0=\Vec{0}$ for the map prior, and so the sky prior must have a mean of 0~K.} at the frequencies in the mixing matrix $S$.

Alternative choices for the component map prior, e.g. using a non-zero $S$ dependent mean, $\mu_0(S)$, and/or an alternative functional form for the covariance matrix, $c_0(S)$, are interesting directions for further work. However, such investigations are beyond the scope of this paper. 

For the remainder of this paper, we will assume that the sky prior covariance matrix is diagonal and that the variance increases at lower frequencies according to a power-law. i.e. that the standard deviation of the sky prior at frequency $v$ is given by the power-law, $A_\mathrm{Sky}\left(v/v_\mathrm{Sky}\right)^{\beta_\mathrm{Sky}}$. This choice of prior for the sky has two parameters, $A_\mathrm{Sky}$ the standard deviation (width) at the chosen reference frequency, and $\beta_\mathrm{Sky}$ the spectral index of the prior. The sky covariance matrix thus has diagonal elements given by:

\begin{equation}
    C_{\mathrm{Sky},{vv}} = A_\mathrm{Sky}^2\left(\frac{v}{v_\mathrm{Sky}}\right)^{2\beta_\mathrm{Sky}}.
\end{equation}

\section{The Synthetic Dataset}
To verify B-GSM's simultaneous component separation and calibration, we conduct studies with a synthetic dataset. For our synthetic dataset, we use two components, which are rescaled according to curved power-law spectra,
\begin{equation}\label{curved powerlaw}
    S^c(v) = \left(\frac{v}{v_0}\right)^{\beta_c + \gamma_c\log\left(v/v_0\right)},
\end{equation}
with $\beta_c$ being the spectral index, $\gamma_c$ being a curvature parameter, and $v_0$ being the reference frequency. The first of these components represents the contribution from the Galactic synchrotron emission, and is scaled using a power-law with $\beta_1=-2.60$ and $\gamma_1=0.0$. The second component represents the contribution from free-free emission, and is scaled according to a power-law with $\beta_2=-2.1$ and $\gamma_2=-0.5$. Figure \ref{f:true comps} shows these components and their spectra.

\subsection{Synthetic Diffuse Dataset: Partial Sky Coverage and Calibration Errors}
These ``true'' components and spectra are used to generate our synthetic dataset. We generate sky maps at  45, 50, 60, 70, 74, 80, 150, 159, and 408~MHz. For each frequency, we rescale the two components using their spectra and sum them (as in equation \ref{eq:component_sep}) to give a realisation of the full synthetic sky at that frequency. 

To reflect the calibration issues present in real diffuse sky maps, we artificially introduce calibration errors into our synthetic diffuse dataset. To do this, we multiply each map in the dataset by a scale factor and apply an offset to the zero level. The known corrections (input values) for these rescalings and offsets are shown in table \ref{t:true cal corrections}, our aim is to recover these calibration corrections during our inference.

\begin{figure}
    \includegraphics[width=0.46\textwidth]{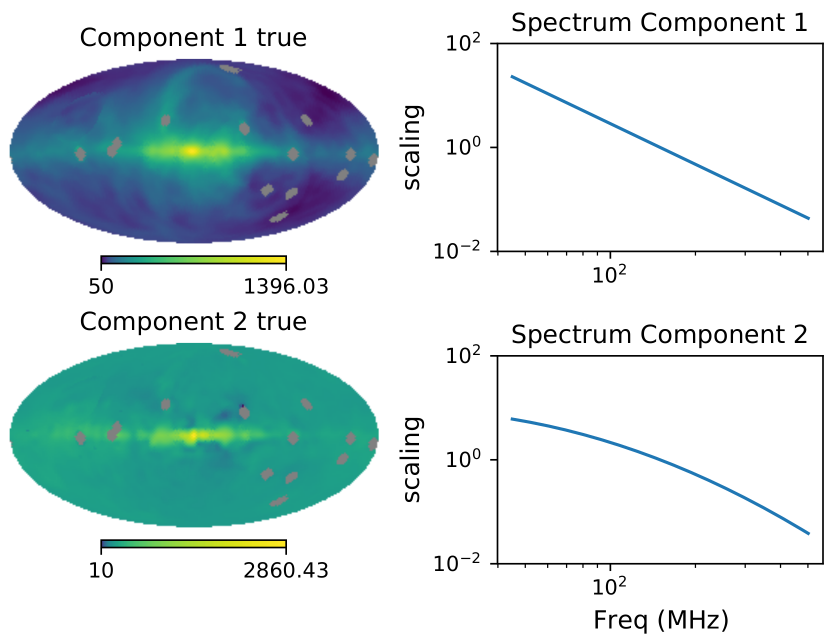}
\caption{The two components and their spectra used to generate the synthetic dataset. The two spectra are power-laws, component 1 has $\beta_1=-2.6$ $\gamma_1=0$, and component 2 has $\beta_2=-2.1$ $\gamma_2=-0.5$. Both component amplitude maps are shown in units of kelvin on a log scale, they are HEALPix maps with $N_\mathrm{side}=32$.\label{f:true comps}}
\begin{center}
    \includegraphics[width=0.48\textwidth]{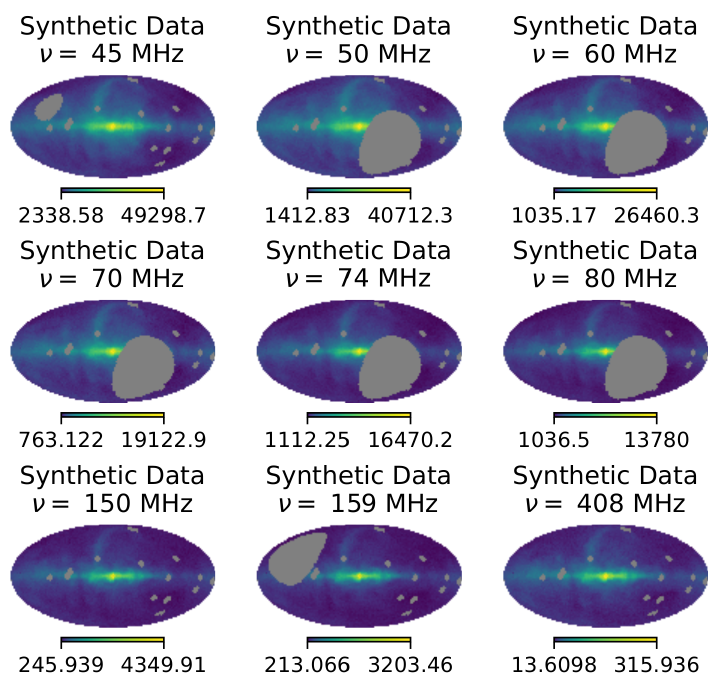}
\end{center}
\caption{The synthetic diffuse dataset after introducing noise and calibration errors.\label{f:synth dataset diffuse}}
\end{figure}

After applying these calibration errors, we add Gaussian noise to the maps with a standard deviation at each pixel equal to 5\% of the map temperature. Additionally, we mask out ``unobserved" sky regions, corresponding to the sky coverage present in the Guzman 45~MHz~\citep{Guzmn2010}, LWA1 maps~\citep{LFSM}, Landecker-Wielebinski (LW) 150~MHz all sky map~\citep{LW150}, EDA2 159~MHz map \citep{EDA2}, and the Haslam 408~MHz all sky map~\citep{Remazeilles2015}. The final synthetic diffuse dataset (shown in figure \ref{f:synth dataset diffuse}) reflects the partial sky coverage, noise, and calibration uncertainty present in real observations of the sky at low frequencies.
\begin{table}
  \medskip
  \centering
    \begin{tabular}{ccc}
      $\nu$ (MHz) & $a_v$ & $b_v$ (K)\\ 
        45 & 1 & 0\\
        50 & 0.95 & +600\\
        60 & 1.05 & +400\\
        70 & 1.06 & +300\\
        74 & 1.10 & -200\\
        80 & 1.15 & -300\\
        150 & 1 & 0\\
        159 & 1.18 & -30\\
        408 & 0.95 & +5\\
    \end{tabular}\\[5pt]
    \caption{The true values of the corrections to the synthetic diffuse dataset calibrations. B-GSM should recover these values for $a_v$ and $b_v$ at each of the frequencies.}\label{t:true cal corrections}
\end{table}

\begin{figure*}
\begin{center}
    \includegraphics[width=0.85\textwidth]{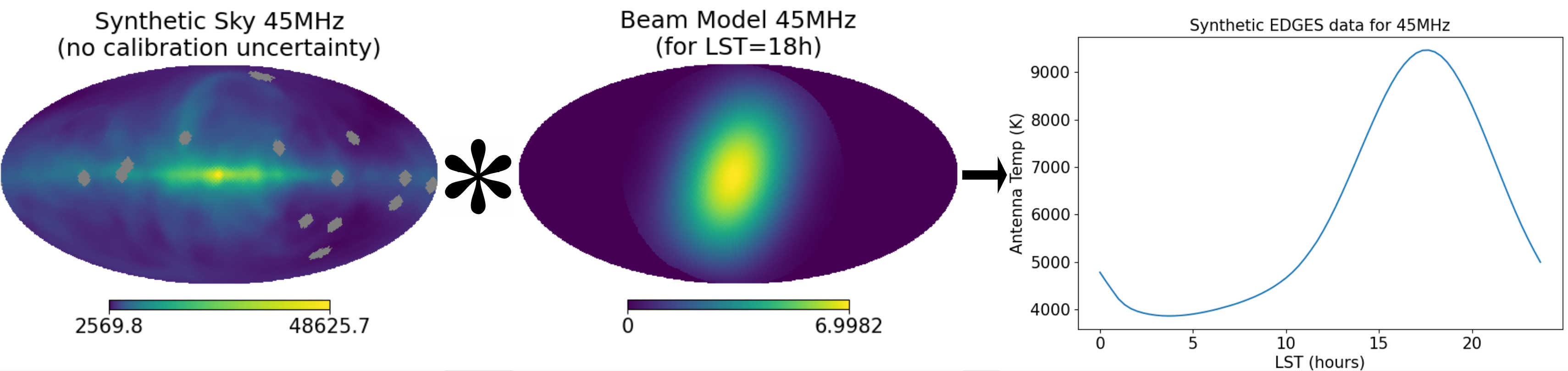}
\end{center}
\caption{Illustration of the production of the synthetic absolute temperature data for 45~MHz. The left-hand panel shows the synthetic full sky map at 45~MHz with no introduction of calibration uncertainty. The centre panel shows the beam model (at LST=18 hours) that this synthetic sky is convolved with, note convolution is performed with beams at every 20 minutes of LST over the range 0 hours to 24 hours. The right-hand panel shows the resulting synthetic $T$ vs LST curve.  \label{f:synth abs dataset}}
\end{figure*}

\subsection{A Synthetic Absolute Temperature Dataset}

To produce the synthetic absolute temperature dataset, we begin by generating synthetic full sky diffuse maps every 5~MHz between 40~MHz and 200~MHz. These are generated using the components and spectra shown in figure \ref{f:true comps}. Note that his frequency range, for the synthetic absolute temperature dataset, was chosen to match the coverage of the EDGES low band and high band~\citep{Mozdzen2018,Mozdzen2017}.

Given these perfectly calibrated synthetic full sky diffuse maps, we produce our synthetic absolute temperature dataset by convolving the map at each frequency with a beam model at that same frequency. We perform this convolution every 20 minutes of LST to produce a perfectly calibrated $T$ vs LST curve at each frequency. We then add Gaussian noise to these $T$ vs LST curves, with a standard deviation equal to 0.2\% of the temperature. This process is illustrated for the synthetic absolute temperature observations at 45~MHz in figure \ref{f:synth abs dataset}. 

Note, for the synthetic absolute dataset, the choice of beam model is arbitrary (as the observations are synthetic, we could choose any beam). We use a Gaussian beam model with a two perpendicular FWHMs. At 45~MHz these FWHMs are $98^{\circ} \times 68^{\circ}$, and at 150~MHz the FWHMs are $112^{\circ}\times 72^{\circ}$, for all other frequencies between we linearly extrapolate the values of the FWHMs at these frequencies. These values very approximately reflect the parameters of the EDGES beam, as reported in \citep{Monsalve2021}.

\section{Results for Synthetic Data}
\subsection{Bayesian Evidence}
\begin{figure}
\begin{center}
    \includegraphics[width=0.48\textwidth]{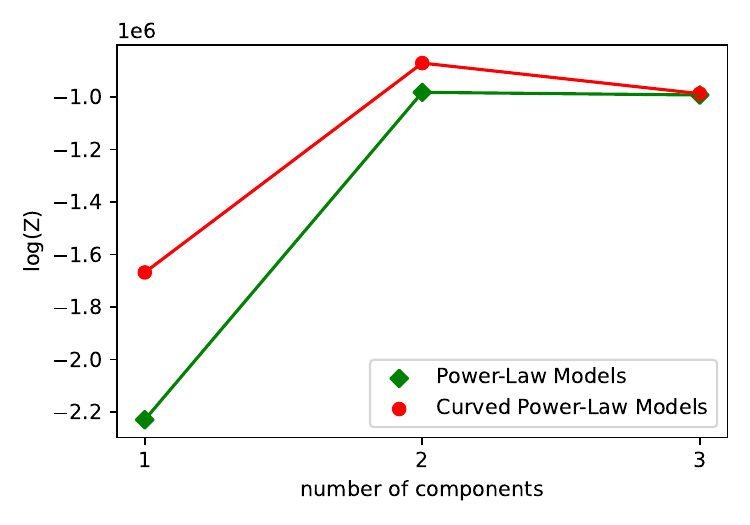}
\end{center}
\caption{Plot of the values found for the Bayesian evidence for each of the tested candidate models. Shown in green are models with component spectra that are power-laws with no curvature ($\gamma_c=0$ $\forall$ $c$). Shown in red are models where all component spectra are curved power-laws. We see that the Bayesian evidence is highest for a 2 component model with curved power-law spectra. We strongly reject the incorrect models, and select for the correct number of components and spectral model type. \label{f:Bayesian evidence syth data}}
\end{figure}
For B-GSM, we want to use the dataset to inform our choice for the number of components and the spectral model for each of these components. We do this using Bayesian model comparison~\citep{Sivia_Skilling}, which allows us to compare the posterior probabilities between models, allowing us to select or reject models informed by the dataset. We tested a range of different candidate models against our synthetic datasets. For each candidate model, we performed nested sampling of the marginal posterior in order to compute a Bayesian evidence value for that model~\citep{Skilling2004,polychord}. We assumed that all candidate models have the same prior probability, meaning that Bayesian model comparison amounts to simply choosing the candidate model with the highest Bayesian evidence.

The flexibility of B-GSM allows us to use any parametrisation for the component spectra. However, for this paper, we tested two different parametrisations for the component spectra. These are:
\begin{itemize}
    \item Power-Law Models: for all components only $\beta$ is a parameter of the spectral model, i.e. for all components we fix $\gamma=0$ (no curved component spectra).
    \item Curved Power-Law Models: all components have curved power-laws, both $\beta$ and $\gamma$ are parameters for the spectral model of all components.
\end{itemize}
For each spectral parametrisation, we tested three different candidate models with: 1, 2, and 3 emission components. In each case, we performed nested sampling using \texttt{polychord} \citep{polychord} with $n_\mathrm{live}=500$ and $n_\mathrm{repeat}$ set to five times the dimensionality of the space. In all cases, the synthetic dataset was identical.

\begin{figure*}
\begin{center}
    \includegraphics[width=0.72\textwidth]{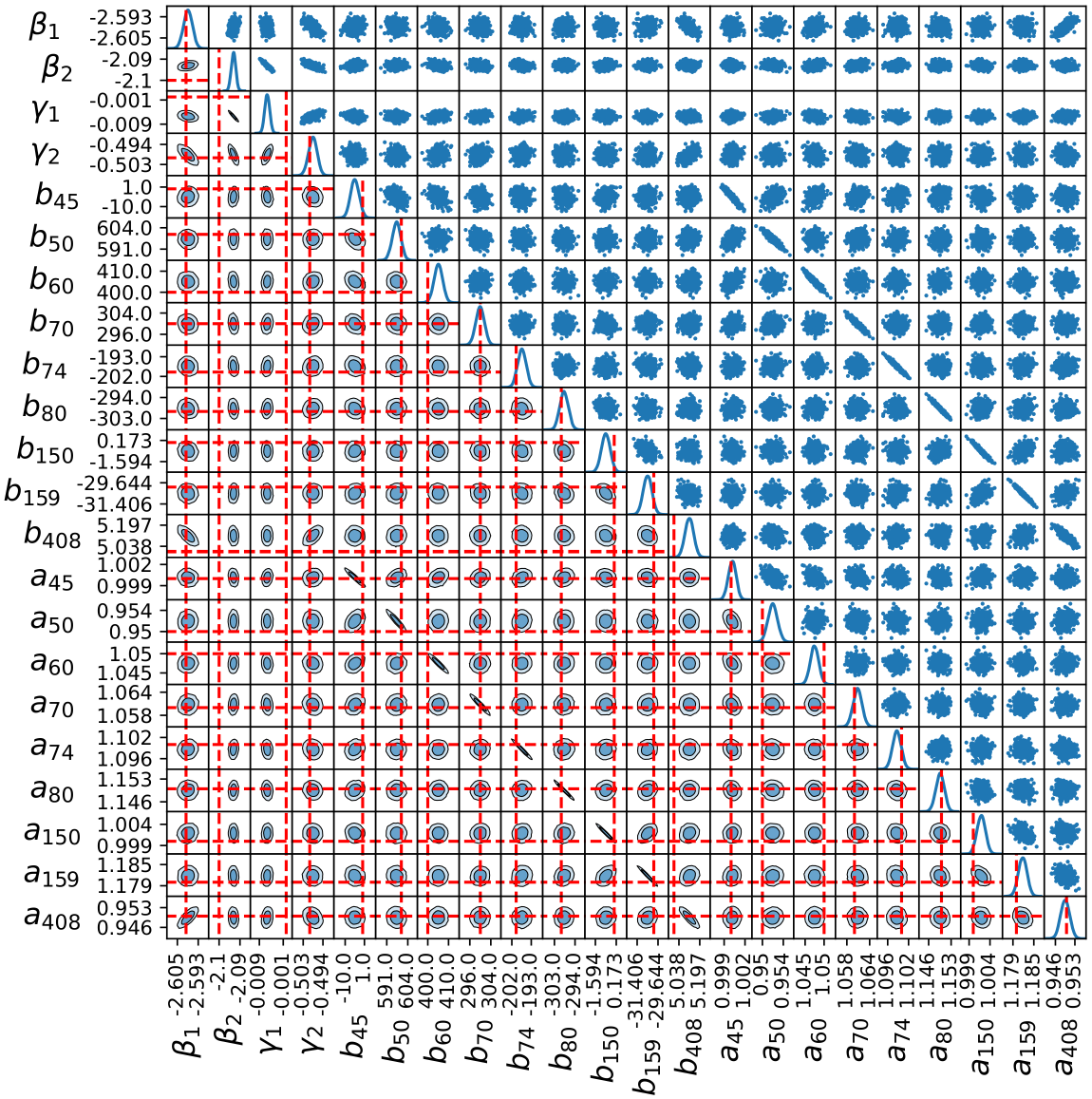}
\end{center}
\caption{Corner plot of the marginal posterior for the highest evidence model, plotted using Anesthetic~\citep{anesthetic}. The contours and points show the marginal posterior distributions for each model parameter. The red dashed lines show the true value (used to generate the synthetic dataset) for each parameter.\label{f:syth posterior samps}}
\end{figure*}

The results we present here use the following priors for the model parameters. For the spectral indexes of the component spectra, we use a uniform prior between -3.5 and +1, $P(\beta_c)=U(-3.5,1)$. For the spectral curvature of the component spectra, we use a Gaussian prior $P(\gamma_c)=N(0,1)$. For the zero level correction, we use a Gaussian prior of width 2000 kelvin at all frequencies $P(b_v)=N(0,2000~K)$. For the temperature scale correction, we use a uniform prior between 0.85 and 1.25 at all frequencies, $P(a_v)=U(0.85,1.25)$. 

As discussed in section \ref{conditional marginalisation}, we must specify a prior covariance for the sky amplitudes, $C_\mathrm{Sky}(A_\mathrm{Sky},\beta_\mathrm{Sky})$. For the results in this section, we take the spectral index to be $\beta_\mathrm{Sky}=-2.8$ and the amplitude to be $A_\mathrm{Sky}=300$~kelvin, with a reference frequency of $v_\mathrm{Sky}=408$~MHz.

We show the Bayesian evidence values computed for each of the six candidate models in figure \ref{f:Bayesian evidence syth data}. The evidence is highest for the candidate model with two emission components and with curved power-law component spectra. This is promising as the synthetic dataset was produced using two components each with a curved power-law spectrum. We see that the incorrect models that can not describe the dataset have significantly lower evidence values and have been strongly rejected. This suggests that B-GSM is able to use the dataset to correctly inform our choice of spectral parametrisation and the number of emission components.

\subsection{Marginal and Component Posteriors}
\begin{figure*}
    \includegraphics[width=0.85\textwidth]{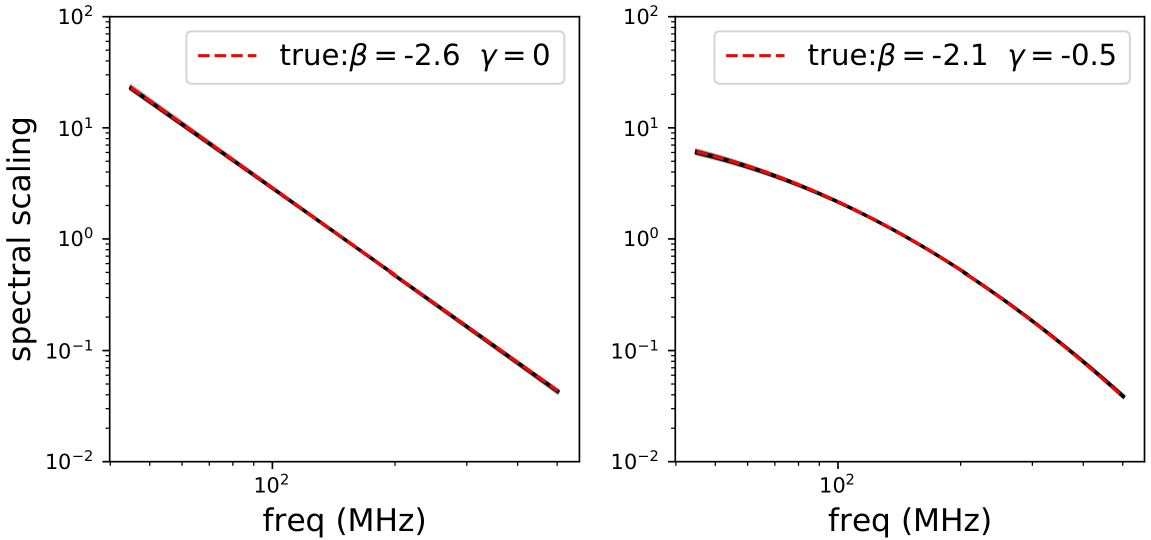}
\caption{Functional posterior plot of component spectra, produced using fgivenx~\citep{fgivenx}. The red dashed lines show the true spectra, used to generate the synthetic dataset. The black lines show posterior spectra, which cluster around the true component spectra, indicating an excellent match to the true spectra across the full frequency range.\label{f:synth spectral post}}
\end{figure*}
Equally important to identifying the correct model, are the values we recover for this model's parameters. To investigate these, we will need to look at the posterior distribution of parameters for our highest evidence model. We show the corner plot of the marginal posterior in figure \ref{f:syth posterior samps}. The contours show the posterior distributions for each model parameter, the red dashed lines show the true value for each model parameter. The first four parameters $\beta_1$, $\beta_2$, $\gamma_1$, $\gamma_2$, are the spectral indexes and curvatures for each of the two component spectra. The following nine parameters correspond to the temperature zero correction for each of the diffuse maps in our dataset, $b_v$. The final nine parameters correspond to the temperature scale correction for each of the diffuse maps in our dataset, $a_v$.

For the spectrum of the first component, we determine a spectral index of $\beta_1=-2.599\pm0.002$ agreeing with the true value of -2.6 within uncertainty, and a curvature of $\gamma_1=0.0064\pm0.0008$ disagreeing with the true value of 0 at a $8\sigma$ level. For the spectrum of the second component, we find a spectral index of $\beta_2=-2.093\pm0.001$ disagreeing with the true value of -2.1 at a $7\sigma$ level, and a curvature of ${\gamma_2=-0.499\pm0.002}$ agreeing with the true value of -0.5 within uncertainty. 

Both the curvature of the first component spectrum, $\gamma_1$, and the spectral index of the second component spectrum, $\beta_2$, disagree with their true values at a $\geq7\sigma$ level, which is concerning. A possible explanation for this discrepancy is that our choice of prior for the component amplitudes is influencing the posterior. The component amplitude prior, $P(\vec{M}_p|S)=N(\mu=0,c_0(S))$, (used during the marginalisation) requires us to specify a prior assumption for the distribution of the diffuse sky map amplitudes at all observed frequencies. For the results in this paper, this sky prior is assumed to be Gaussian with a mean of 0 K with a diagonal covariance matrix, $C_\mathrm{Sky}$. The diagonal elements of this sky covariance matrix, $C_\mathrm{Sky}$, are assumed to follow a power-law, with the variance increasing at lower frequencies. The choice of spectral index for this sky prior covariance matrix, $\beta_\mathrm{Sky}$, is likely to impact the posterior distribution of our model parameters. We investigate and attempt to quantify the biasing effect of our choice of, $\beta_\mathrm{Sky}$, in section \ref{impact of prior section} of this paper.  

We note that for both $\gamma_1$ and $\beta_2$, the discrepancy between the true and posterior mean values, is small in absolute terms. In figure \ref{f:synth spectral post} we show the functional posterior for the two component spectra. The black lines are the posterior spectra\footnote{Note that for each component, we plot a posterior spectrum (black line) for each of the $\sim30000$ posterior samples of the parameters $\beta_c$ and $\gamma_c$. Due to the fact that the posterior samples of these parameters all have similar values, these $\sim 30000$ spectra cluster together and look like a single line for each component.}, the red lines are the true spectrum for each component. We see that the posterior spectra cluster around the true spectra, and that across the full frequency range the posterior spectra are in excellent agreement with the true spectra. It appears that the discrepancy between the true and posterior mean values for the parameters $\gamma_1$ and $\beta_2$ does not significantly impact the component spectra across the narrow frequency range of B-GSM. 

There are correlations between the posterior distributions of parameters of the spectral model and the calibration at 408~MHz. The spectral index $\beta_1$ shows a strong positive correlation with the scale correction, $a_{408}$, and a strong negative correlation with the zero correction, $b_{408}$, at 408~MHz. The spectral curvature of the second component, $\beta_2$, shows a strong positive correlation with the zero correction, $b_{408}$. We also note that the spectral parameters show correlations. We see that $\beta_1$ shows a strong negative correlation with $\gamma_2$, and that $\gamma_1$ shows a strong negative correlation with $\beta_2$. 

We see that the posterior distributions of the calibration parameters, $a_v$ and $b_v$, for the maps at 45, 50, 60, 70, 74, 80, 150, and 159~MHz agree with the true values at an approximately $2\sigma$ level. Indicating that B-GSM has correctly determined calibration for these maps.

For the map at 408~MHz, the posterior temperature scale correction, $a_{408}=0.949\pm0.002$, agrees with the true value within uncertainty. However, the posterior mean zero level correction, $b_{408}=5.12\pm0.03$ K, disagrees with the true value of 5 K at a $\sim 4\sigma$ level. We note that the discrepancy from the true value is small in absolute terms, and that B-GSM has identified a correction to the zero level that is approximately correct within roughly 0.1~K. This is despite our very broad prior assumption of ${P(b_{408})=N(\mu=0~\mathrm{K},\sigma=2000~\mathrm{K})}$.

A possible explanation for the discrepancy between the posterior zero correction at 408~MHz and the true value is that we do not have any direct measurements of the sky absolute temperature for any frequency above 200~MHz. Calibration of the 408~MHz map is done by extrapolating the absolute temperature data, from lower frequencies, using the spectral model and sky predictions of B-GSM. We have imperfectly recovered our spectral model, it is therefore unsurprising that we have imperfectly calibrated the map at 408~MHz. Additionally, the fact that the calibration of the 408~MHz map relies on B-GSMs sky predictions and spectral model provides an explanation for the correlations we see between the spectral model parameters, $\beta_1$ and $\gamma_2$, and the 408~MHz calibration parameters, $a_{408}$ and $b_{408}$.

\begin{figure}
\begin{center}
    \includegraphics[width=0.48\textwidth]{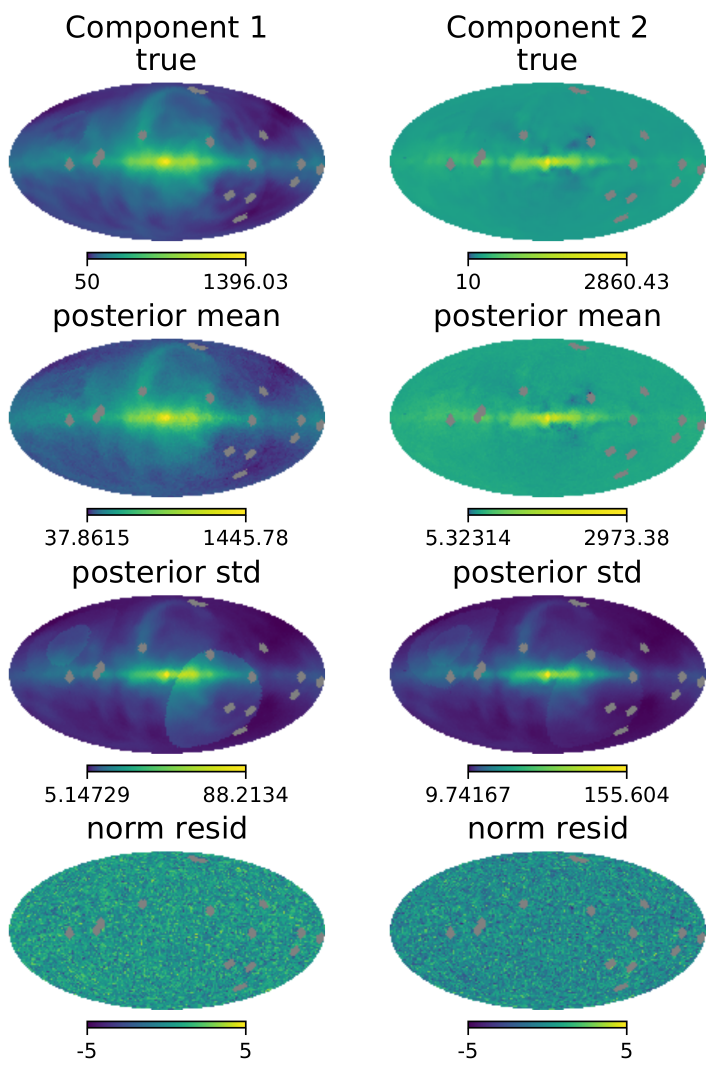}
\end{center}
\caption{Posterior mean for the component amplitude maps. The top row shows the true component amplitude maps used to generate the synthetic dataset. The second row shows the weighted mean of the posterior samples for the component amplitudes, the third row shows the weighted standard deviation of component amplitudes. The bottom row shows the normalised residuals between the true amplitudes and the posterior mean amplitudes. The normalised residuals are white noise.\label{f:synth comp post}}
\end{figure}

\begin{figure}
\begin{center}
    \includegraphics[width=0.48\textwidth]{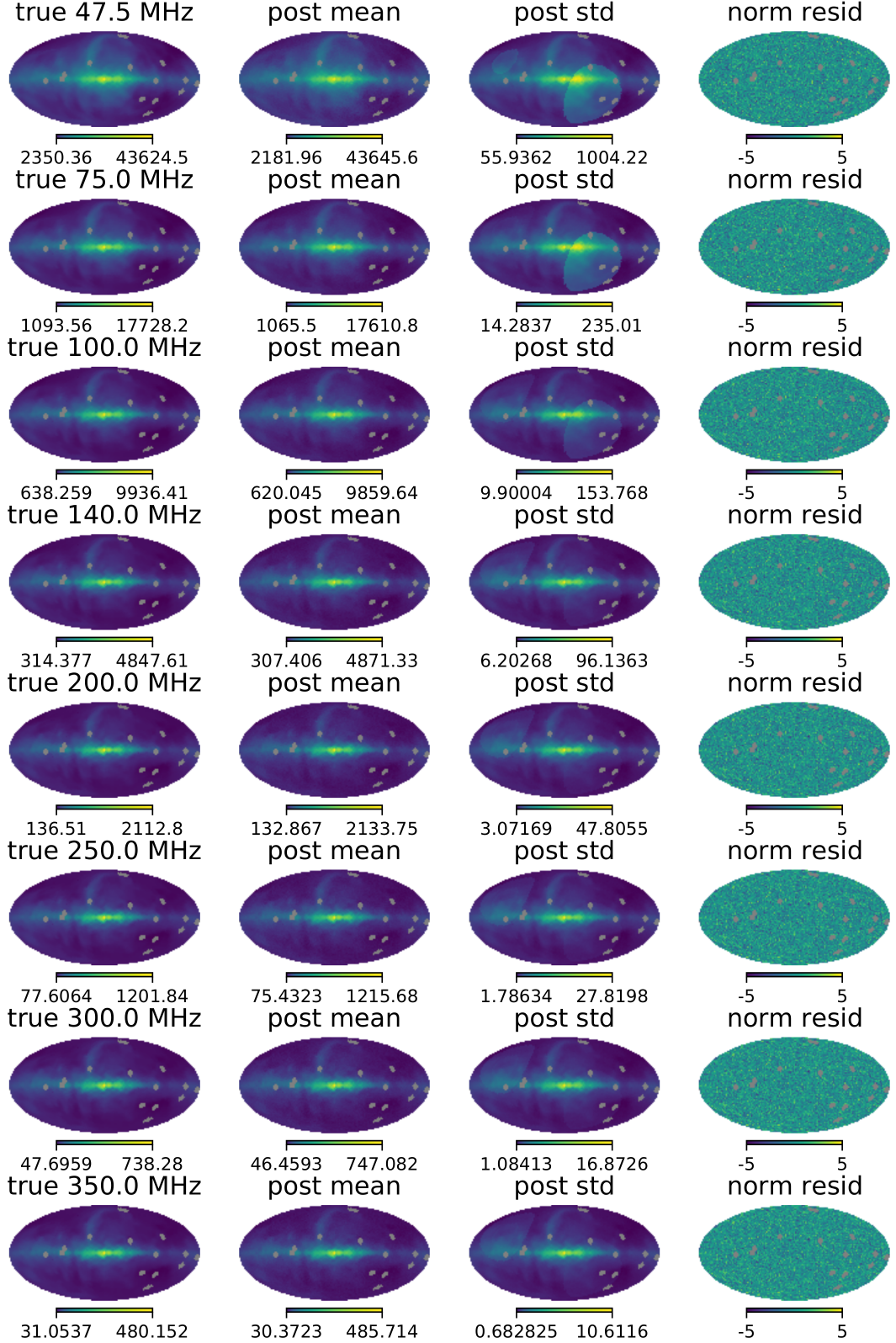}
\end{center}
\caption{The posterior distribution for the sky at a selection of frequencies covering the full frequency range of B-GSM. We can see that the recovery of the sky is excellent at all frequencies, with the normalised residuals being white noise.\label{f:synth sky post}}
\end{figure}

So far we have only discussed the marginal posterior of the model parameters. However, we must also discuss the posterior distribution of the component amplitudes. We generate samples from the posterior distribution of component amplitudes, $\{M\}_\mathrm{posterior}$. To do this, we generate a sample set of component amplitude maps for each of the samples in the marginal posterior. The $i$-th posterior sample amplitude maps, $M_i$, is drawn from the conditional distribution, $M_i \sim P(M|a_i,\Vec{b}_i,S_i, D)$, conditioned on the $i$-th sample from the marginal posterior $a_i,\Vec{b}_i,S_i$ and our observed data. The mean posterior component amplitude maps are shown in figure \ref{f:synth comp post}.

We see that the posterior mean component maps are indistinguishable by eye from the true component maps. Looking at the third row, we see the standard deviation of the posterior component map samples, effectively the statistical uncertainty on the posterior mean. The posterior standard deviation is larger in the poorly observed southern sky and regions where the sky is brightest. In the fourth row, we show the normalised residuals, (posterior - true)/$\sigma_\mathrm{posterior}$. The normalised residuals are white noise for both components, indicating that the component amplitudes are recovered within statistical uncertainty.
\begin{table}
  \centering
    \begin{tabular}{ccc}
        $v$ (MHz) & RMS Residual (kelvin):  & RMS Residual (kelvin) \\ 
         & (Uncalibrated dataset) & (Posterior Mean) \\\hline
         45 & 18.14 & 12.86\\
         50 & 426.61 & 11.79\\
         60 & 563.17 & 7.81\\
         70 & 437.29 & 4.54\\
         74 & 51.17 & 5.11\\
         80 & 74.09 & 4.31\\
         150 & 1.16 & 0.73\\
         159 & 35.20 & 0.69\\
         408 & 3.78 & 0.12
    \end{tabular}\\[0pt]
    \caption{RMS residual at each frequency, between the $T$ vs LST curves of the true signal and both the uncalibrated dataset (column 1) and posterior mean (column 2).}\label{t:syth TvsLST RMS}
\end{table}

\subsection{Posterior Sky Predictions}
At this point we have generated samples drawn from the posterior distribution of the spectra, $\{S\}_\mathrm{posterior}$, and the posterior of component maps, $\{M\}_\mathrm{posterior}$. We use these to determine a prediction for the sky and its uncertainty at any given frequency, $v$. To do this, we generate a set of posterior sample predictions for the sky at frequency $v$, $\{\mathrm{Sky}_{i,v}\}_\mathrm{posterior}$. Each posterior sample prediction of the sky, $\mathrm{Sky}_{i,v}$, is calculated using equation \ref{eq:component_sep} for the sample spectral parameters, $S_i$, and sample component amplitude maps, $M_i$. The mean prediction for the sky map at frequency $v$ is given by the weighted mean of these posterior samples, and the uncertainty map is the weighted standard deviation of these posterior samples.

In figure \ref{f:synth sky post} we compare these posterior sky predictions to the true sky, at a selection of frequencies covering the full frequency range of B-GSM. The first column shows the true sky at each frequency, these maps are generated using the true component amplitude maps and spectra (shown in figure \ref{f:true comps}), they have no noise or calibration uncertainty (by definition they are the true underlying signal at each frequency). The second column shows the posterior mean sky prediction at each frequency, and the third column shows the posterior standard deviation at each frequency. We see that at all frequencies the spatial distribution and amplitude for the posterior sky is in excellent agreement with the true underlying signal. Additionally, we see that the uncertainty is largest in the poorly observed southern sky at low frequencies, indicating that as expected the posterior distribution is broader in the regions with limited data.

In the fourth column of figure \ref{f:synth sky post} we show (for each frequency) maps of the spatial distribution of normalised residuals between the true underlying signal and the posterior mean prediction. We see that for all frequencies tested, these normalised residuals show no structure, indicating that the posterior sky predictions agree with the true signal, within statistical uncertainty, across the whole sky for the full frequency range.

\begin{figure}
\begin{center}
    \includegraphics[width=0.48\textwidth]{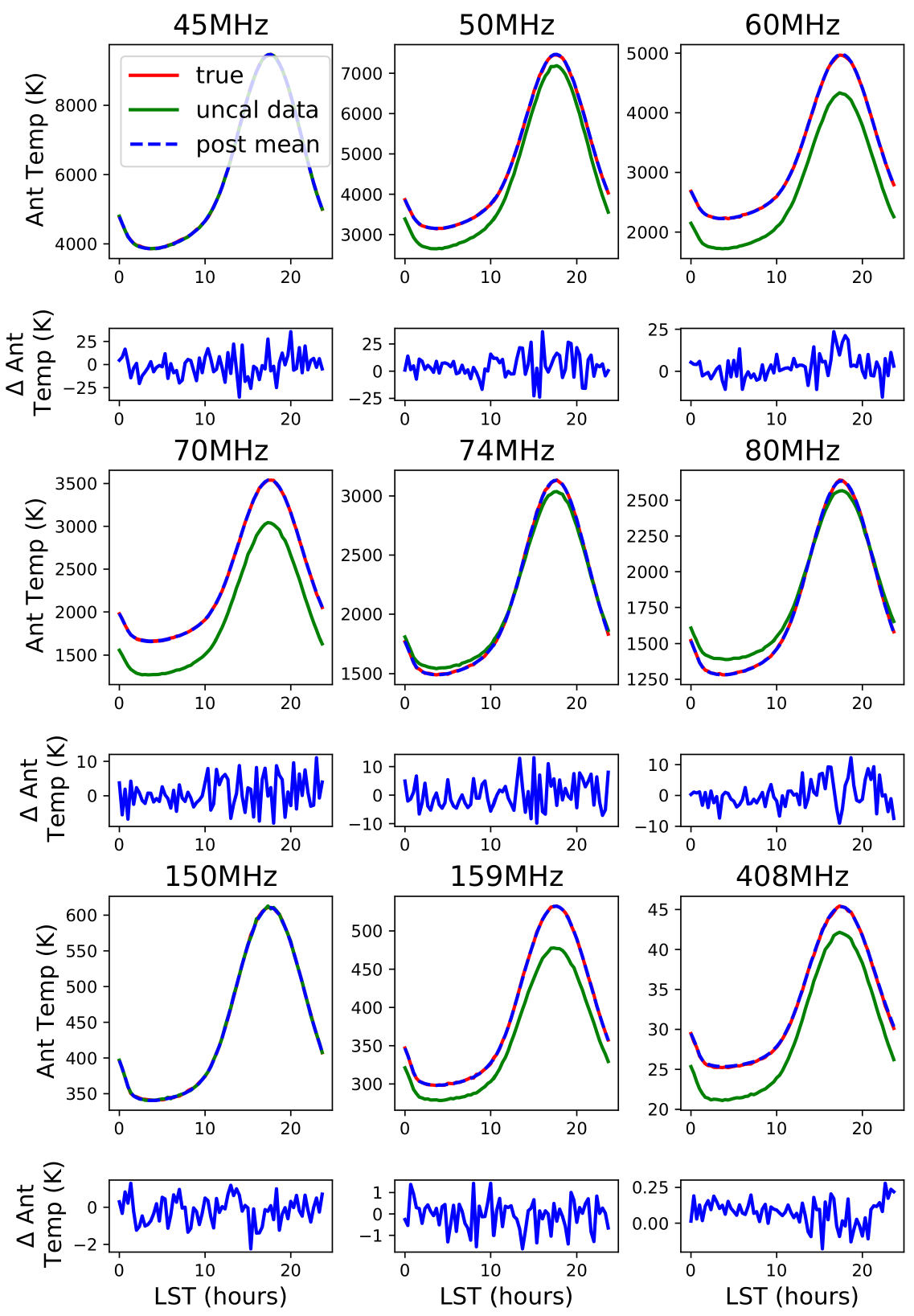}
\end{center}
\caption{Comparison of the $T$ vs LST curves for the true underlying signal (red curves), mean posterior sky prediction (blue dashed curves), and uncalibrated synthetic diffuse dataset (green curves). The smaller panels show the residual between the mean posterior prediction and the true underlying signal (posterior - true) for each frequency. Across the full frequency range, the model posterior is in excellent agreement with the true underlying signal. \label{f:synth T vs LST comparison}}
\end{figure}

In addition to comparing the posterior to the diffuse sky, we also compare the posterior to the absolute temperature measurements. This is shown is figure \ref{f:synth T vs LST comparison} where we compare the true $T$ vs LST curves (red curves), with both the uncalibrated dataset (green curves), and the posterior mean predictions (blue curves). The larger panels show the $T$ vs LST curves at each frequency, and the smaller panels show the residual between the true signal and the posterior predictions (defined as posterior - true) for each frequency. The posterior $T$ vs LST curves are in excellent agreement with the true $T$ vs LST for all tested frequencies, and the residuals at all frequencies have a noise like distribution with a mean of $\sim$0~K.

In table \ref{t:syth TvsLST RMS} we show the root-mean-square (RMS) residual (in K) for the $T$ vs LST curves of the uncalibrated dataset and the posterior mean predictions. We see a significant reduction in RMS for the posterior predictions, when compared with the uncalibrated dataset. This reduction in RMS and the excellent match between the posterior and true $T$ vs LST curves, shown in figure \ref{f:synth T vs LST comparison}, indicates that the posterior sky predictions for B-GSM have been successfully calibrated to the absolute temperature dataset.

\section{Impact of Choice of Prior}\label{impact of prior section}
As discussed in the previous section, the choice of prior for the component amplitude maps is likely to bias the posterior distributions of parameters for B-GSM. This is due to our conditional component amplitude prior, $P(\Vec{M}_p|S)$, requiring us to specify a prior for the sky amplitudes, which we assume to be Gaussian with a mean of 0~K at all observed frequencies and with a diagonal covariance matrix $C_\mathrm{Sky}(A_\mathrm{Sky},\beta_\mathrm{Sky})$. We want to quantify the impact of varying the sky prior covariance matrix parameters, $A_\mathrm{Sky}$ and $\beta_\mathrm{Sky}$, on the posterior distribution for B-GSM. To do this, we sampled the posterior for a two component model (with curved power-law component spectra) for a selection of different values of $A_\mathrm{Sky}$ and $\beta_\mathrm{Sky}$, all other priors and the dataset are kept the same between sampling runs. We found that the choice of the prior covariance width, $A_\mathrm{Sky}$, has no statistically significant impact on the posterior distributions for any model parameters or on the final posterior sky predictions. However, changes to the prior covariance spectral index, $\beta_\mathrm{Sky}$, do have a statistically significant impact on the posterior distributions of the model parameters.

\begin{figure}
\begin{center}
    \includegraphics[width=0.48\textwidth]{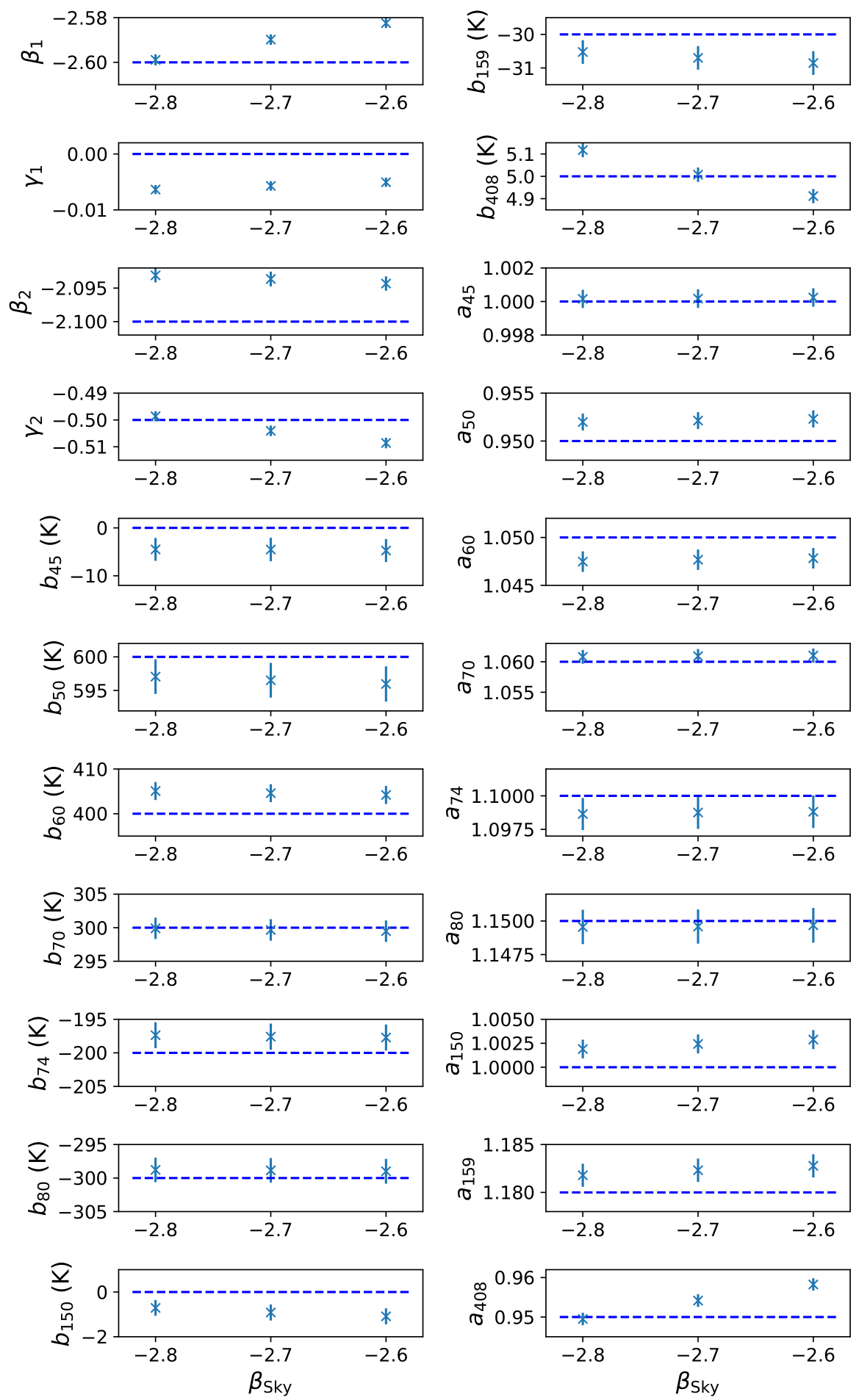}
\end{center}
\caption{Comparison of mean posterior parameters for a selection of different priors, the x-axis is the value for the spectral index of the sky prior covariance matrix, $\beta_\mathrm{Sky}$, used for each run. The coloured dots show the mean posterior value for each parameter for each of the tested priors, the errorbars show the $1\sigma$ uncertainty. The horizontal blue dashed lines show the true value (used to generate the synthetic dataset) for each parameter. \label{f:different priors}}
\end{figure}

In figure \ref{f:different priors}, we summarise the marginal posterior (of the spectral and calibration parameters) for three different choices of the spectral index of the sky prior covariance; $\beta_\mathrm{Sky}=-2.6$, $\beta_\mathrm{Sky}=-2.7$, and $\beta_\mathrm{Sky}=-2.8$. Each panel in the figure corresponds to a specific model parameter, the crosses show the posterior mean value (for each of the three tested priors) and the errorbar shows the $1\sigma$ uncertainty, the blue dashed line shows the true value for the parameter. We see that variation of $\beta_\mathrm{Sky}$ results in pronounced changes to the posterior mean values of the spectral parameters ($\beta_1$, $\beta_2$, $\gamma_1$ and $\gamma_2$) and the 408~MHz calibration parameters ($a_{408}$ $b_{408}$).

We see that both the spectral index of the first component, $\beta_1$, and the temperature scale correction at 408~MHz, $a_{408}$, show a strong positive correlation with the choice of $\beta_\mathrm{Sky}$. I.E. selecting a less negative (less steep) prior spectral index $\beta_\mathrm{Sky}$ results in a less steep posterior spectral index for the first component and yields a higher value for the scale correction of the map at 408~MHz. Additionally, the correction to the zero level for the map at 408~MHz, $b_{408}$, and the spectral curvature of the second component, $\gamma_2$, both display negative correlations with the prior choice of $\beta_\mathrm{sky}$. We note that, the posterior calibration parameters, $a_v$ and $b_v$, for the maps below 200~MHz (where we have absolute temperature data to act as a ground truth) show little dependence on the prior choice of $\beta_\mathrm{Sky}$.
\begin{figure}
\begin{center}
    \includegraphics[width=0.48\textwidth]{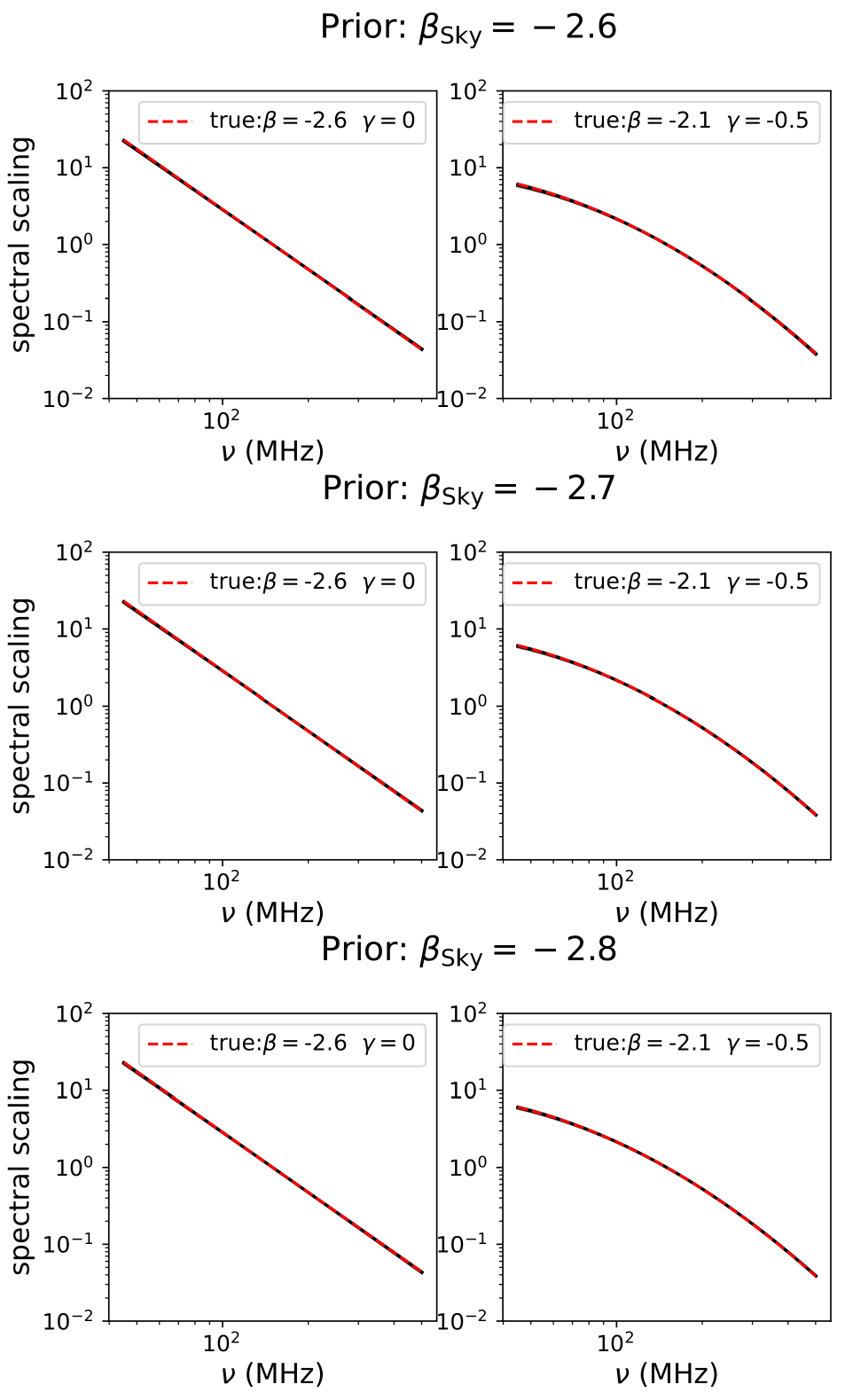}
\end{center}
\caption{The functional posterior of the component spectra for the three alternate priors. Shown here are the posteriors for priors with $\beta_\mathrm{Sky}=-2.8$, $-2.7$, and $-2.6$. We note that despite the statistically significant change to the posterior spectral parameters, the functional forms remain an excellent match to the true spectra.\label{f:different priors spectral posterior}}
\end{figure}
\begin{figure}
\begin{center}
    \includegraphics[width=0.45\textwidth]{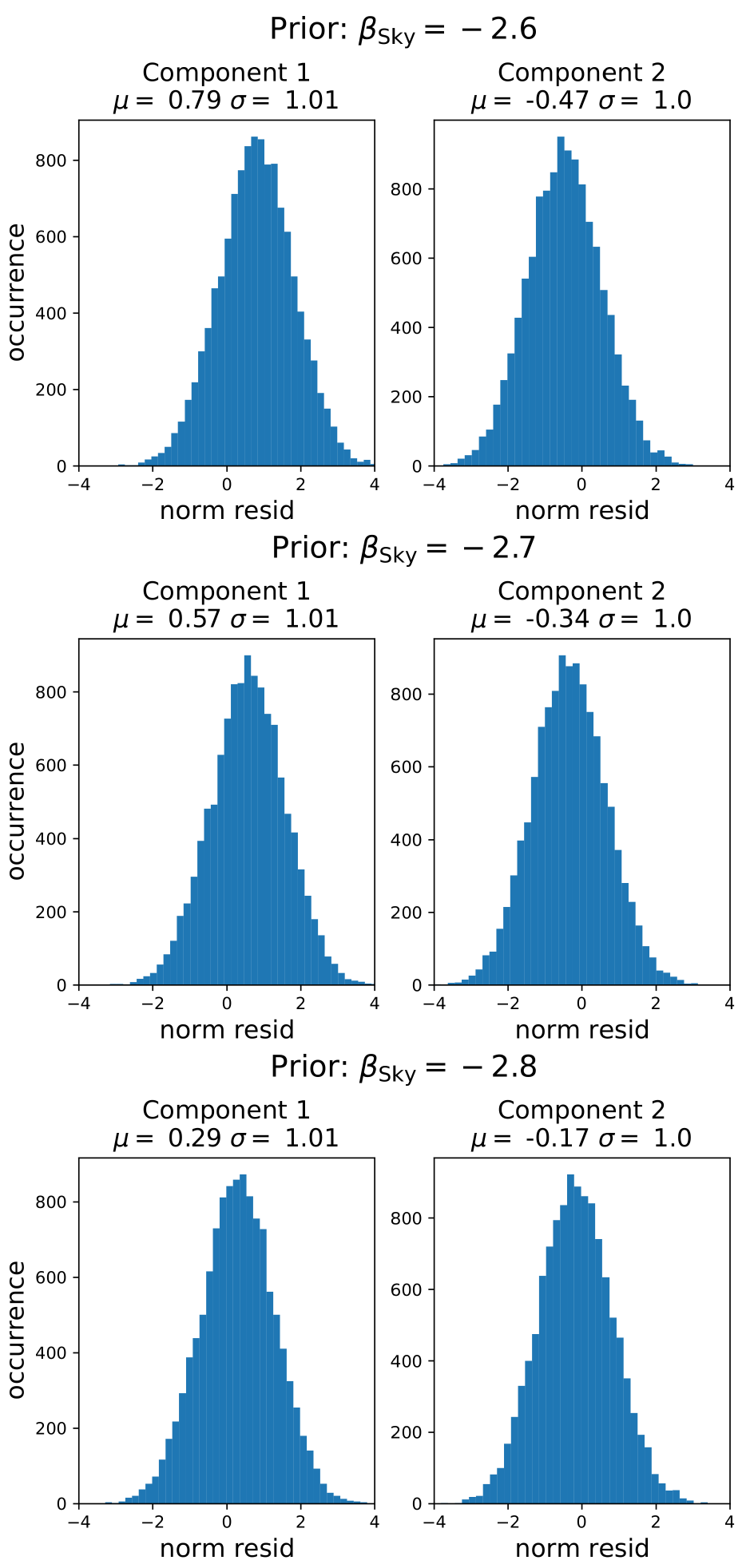}
\end{center}
\caption{Histograms of the normalised residuals between the mean posterior and the true component amplitude maps, $~{(\left<\mathrm{M}\right>_\mathrm{posterior} - \mathrm{true})/\sigma_\mathrm{posterior}}$. Each row shows the histograms for each of the two components (1st component on the left, 2nd on the right), for each of the tested alternative sky prior covariance matrices. We see that for all tested priors, the histograms for both components are Gaussian with standard deviation $\sim 1$, indicating that the residuals are noise-like. However, we see that the mean of the distributions changes as we alter the prior.\label{comp histo for all tested priors}}
\end{figure}
\begin{figure*}
\begin{center}
    \includegraphics[width=0.98\textwidth]{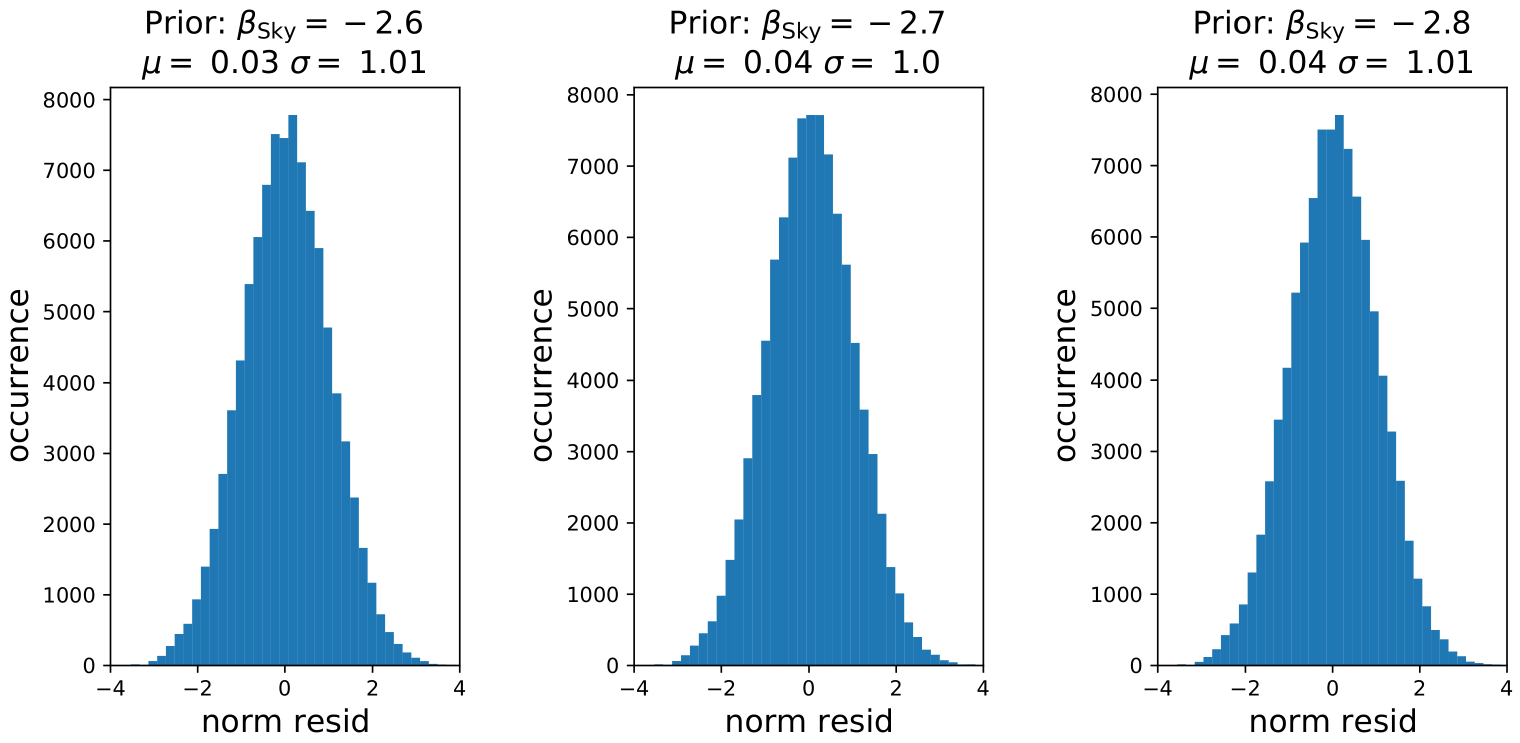}
\end{center}
\caption{Histograms of normalised residuals between mean posterior sky prediction and the true sky, $(\left<\mathrm{Sky}\right>_\mathrm{posterior} - \mathrm{true})/\sigma_\mathrm{posterior}$, for each of the tested alternative sky prior covariance matrices. Note the histograms are for the combined normalised residuals for comparisons at 47.5, 75, 100, 140, 200, 250, 300, and 350~MHz i.e. across the full frequency range of B-GSM. We can see that for all tested priors, the normalised residuals are Gaussian distributed with a mean of $\sim0$ and standard deviation $\sim1$. Indicating that for all tested priors, the normalised residuals are white noise and that in all cases the posterior predictions for the sky agree within uncertainty with the true sky. \label{histo for all tested priors}}
\end{figure*}

The variations in the posterior mean values, due to the choice of prior, are small in absolute terms. Indeed, if we look at the functional posterior plots for the component spectra, shown in figure \ref{f:different priors spectral posterior}, we see that for all three tested prior values of $\beta_\mathrm{Sky}$ the true and posterior spectra are indistinguishable by eye. This suggests that we are able to approximately recover the correct spectral behaviour for the emission components, even after altering our prior assumption for $\beta_\mathrm{Sky}$. Nonetheless, we must acknowledge that variation of the prior sky covariance matrix, $C_\mathrm{Sky}(A_\mathrm{Sky},\beta_\mathrm{Sky})$, does have a statistically significant impact on the marginal posterior.

In addition to looking at how the prior influences the marginal posterior, we investigated how the choice of $C_\mathrm{Sky}$ influences the posterior component amplitudes for B-GSM. We find that for all three tested priors we, the resulting mean posterior component amplitude maps closely matched the spatial structure and amplitudes of the true components, i.e. for all three priors the maps of the estimated and true component amplitudes are visually indistinguishable. Suggesting that B-GSM accurately identifies the spatial distribution of emission components for a range of different prior values of $\beta_\mathrm{Sky}$.

However, if we look at the distribution of the normalised residuals, we note that there is a statistically significant change in the distribution in response to variation of $\beta_\mathrm{Sky}$. This can be seen in the normalised residual histograms for the component maps (figure \ref{comp histo for all tested priors}). We see that the mean value of the distribution of normalised residuals changes for the priors with different values of $\beta_\mathrm{Sky}$. That is to say, as we change our prior, the discrepancy between the true component amplitude maps and the posterior mean component amplitudes changes in response. A possible explanation for this is that our choice of prior impacts our posterior spectra, and the component amplitudes must change in response to these changing spectra.

We must note that while the posterior of both the spectral parameters and component amplitudes change noticeably, as we vary $\beta_\mathrm{Sky}$, the posterior distribution of sky predictions changes to a far lesser degree. The RMS residuals between the true sky $T$ vs LST and posterior $T$ vs LST show only minor changes between the three tested priors, and in all cases we find a significant reduction in RMS for the posterior predictions when compared to the uncalibrated dataset. Indicating that in all three cases, B-GSM is able to calibrate its posterior sky predictions to the absolute temperature dataset. 

Likewise, for all three tested priors, we find that the posterior predicted sky is in excellent agreement with the true sky across the full frequency range of B-GSM. This is shown in the distributions of normalised residuals between the true sky and posterior mean sky for each of the three priors (shown in figure \ref{histo for all tested priors}). We see that for all three priors, the normalised residuals follow a Gaussian distribution with a mean of $\sim0$ and standard deviation of $\sim1$, indicating that the posterior predicted sky and the true sky agree within uncertainty with noise-like residuals. This suggests that the posterior sky predictions of B-GSM show minimal dependence on our choice for the prior covariance matrix of the sky, $C_\mathrm{Sky}(\beta_\mathrm{Sky},A_\mathrm{Sky})$. 

\section{Conclusions}
In this paper, we have demonstrated the Bayesian Global Sky Model (B-GSM) on a synthetic dataset. Bayesian model comparison was used to determine the optimal number of emission components and their spectral parametrisation, achieving the highest Bayesian evidence for a two-component model with curved power-law spectra. This demonstrates B-GSMs ability to use the dataset to reject unsuitable models, and correctly identify the appropriate number of emission components and spectral parameterisation. 

For this highest evidence model, the posterior sky predictions are found to agree within uncertainty with the true synthetic sky (figure \ref{f:synth sky post}), and the posterior predictions of Temperature as a function of LST ($T$ vs LST) are found to be in excellent agreement with their true values at all tested frequencies (figure \ref{f:synth T vs LST comparison}). Additionally, we found a significant reduction in RMS residuals when comparing the $T$ vs LST curves of the uncalibrated dataset and the posterior predictions, demonstrating B-GSMs ability to calibrate its posterior sky predictions to the independent absolute temperature dataset. This indicates that B-GSM is able to successfully integrate both the diffuse and absolute temperature datasets into its inference.

We found that the posterior distributions for B-GSMs component spectra, component amplitudes, and calibration at 408~MHz change (to a statistically significant extent) in response to variation of the prior sky covariance matrix's spectral index parameter $\beta_\mathrm{Sky}$. The variations are small in absolute terms. For all tested priors, the functional plots of the posterior component spectra (Figure \ref{f:different priors spectral posterior}) show excellent agreement with the true spectra. Additionally, the posterior mean component amplitudes share the same spatial structure and have similar amplitudes (within a few kelvin) as the true component amplitudes.  

The posterior calibration parameters for the diffuse maps at frequencies $<200$~MHz (where we have absolute temperature data) show negligible variation between the three different priors. The posterior calibration at 408~MHz varies as we change the value of $\beta_\mathrm{Sky}$. However, for all tested priors and the posterior calibration, $a_{408}$ and $b_{408}$, is within a few percent of the true value. This suggests that while the choice of the prior value of $\beta_\mathrm{Sky}$ does impact the posterior calibration for the map at 408~MHz, we still approximately recover correct the calibration correction. 

Crucially, we find that despite these changes to the posterior spectra and component amplitudes, all tested priors result in posterior sky predictions that agree within uncertainty with the true sky across our full frequency range. Indicating that B-GSM is able to robustly model the true underlying sky within statistical uncertainty. This robustness ensures that B-GSM will be able to reliably and accurately model the low frequency sky during its deployment on a real dataset. 

Further investigation of the prior sensitivity of B-GSMs posterior is necessary, in particular we will investigate an alternative synthetic dataset to see how the choice of data (which frequencies are observed) influences the posterior. Investigations of using more general spectra in B-GSM for example non-parametric models such as FlexKnot~\citep{Shen2024}, and broken power-laws would allow us to fully demonstrate the flexibility of the B-GSM framework. In the second paper of this series, we will apply B-GSM to real observational data to produce a new model of the galactic diffuse radio foreground at frequencies below 400~MHz. This foreground model will be publicly released and will be available for download.

%\section*{Acknowledgements}

%The Acknowledgements section is not numbered. Here you can thank helpful
%colleagues, acknowledge funding agencies, telescopes and facilities used etc.

%%%%%%%%%%%%%%%%%%%%%%%%%%%%%%%%%%%%%%%%%%%%%%%%%%
\section*{Data Availability}
The synthetic dataset used in this study along with the code for B-GSM are available for public download from the following GitHub repository:\newline \url{https://github.com/George-GTC30/Bayesian-Global-Sky-Model-B-GSM-Paper-1}


\newpage
\bibliographystyle{mnras}
\bibliography{example} % if your bibtex file is called example.bib

%\begin{thebibliography}{99}
%\end{thebibliography}

%%%%%%%%%%%%%%%%%%%%%%%%%%%%%%%%%%%%%%%%%%%%%%%%%%

%%%%%%%%%%%%%%%%% APPENDICES %%%%%%%%%%%%%%%%%%%%%
\appendix
\section{Validation of the Approximation}
\begin{figure}\label{histo of abs temp likelihoods}
    \centering
    \includegraphics[width=0.99\linewidth]{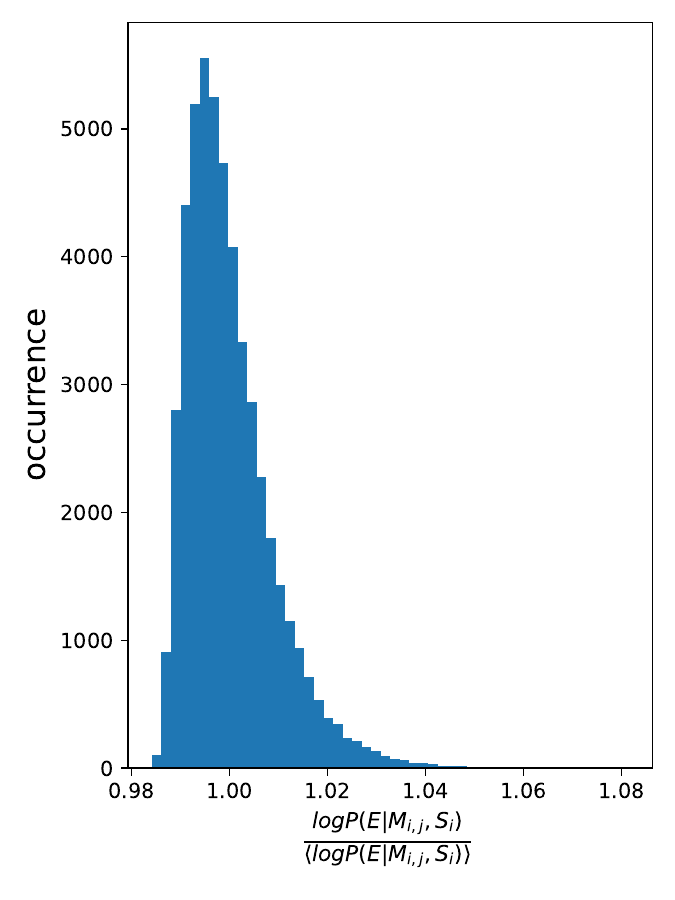}
    \caption{Histogram of values of the absolute temperature likelihood, $P(E|\Vec{M},S)$ evaluated at 50,000 points in the diffuse joint posterior $P(a,\Vec{b},\Vec{M},S|D)$. We see that all 50,000 values of $log P(E|M,S)$ are similar, and that $79.8\%$ are within $1\%$ of the mean value. This suggests that the simplification we use in our approximate marginal likelihood is reasonable.}
    \label{fig:enter-label}
\end{figure}
To validate the assumption at the core of our approximate likelihood we use the 5000 highest likelihood marginal posterior samples (from our two component model with curved power-law spectra, marginal posterior shown in figure \ref{f:syth posterior samps}). For each of these samples, $a_i,\Vec{b}_i,S_i$, we draw 10 examples of component amplitude maps from the conditional distribution $\Vec{M}_{i,j}\sim P(\Vec{M}_i|a_i,\Vec{b}_i,S_i,D)$, and compute the value of the absolute temperature likelihood for each of these 10 example component amplitude maps $P(E|M_{i,j},S_i)$. This gives us 50,000 different values of the absolute temperature likelihood, evaluated at 50,000 different points covering a wide region of the joint diffuse posterior $P(a,\Vec{b},\Vec{M},S|D)$. If our approximation is true then these 50,000 absolute temperature values should all be similar. 

In figure \ref{histo of abs temp likelihoods} we show the histogram of these 50,000 absolute temperature likelihood values. We find that 79.8\% of the values are within 1\% of the mean value, and that for 96.5\% of the posterior points the log absolute temperature likelihood is within 2\% of the mean value. This variation in the log absolute temperature likelihood is small relative to the difference in evidence between different model parametrisations. Indicating that our choice to use the conditional mean set of component amplitude maps, $\langle\Vec{M}|a,\Vec{b},S,D\rangle$, has minimal impact on the log likelihood. This suggests that our approximate marginal likelihood is reasonable. 
%%%%%%%%%%%%%%%%%%%%%%%%%%%%%%%%%%%%%%%%%%%%%%%%%%

% Don't change these lines
\bsp	% typesetting comment
\label{lastpage}
\end{document}